\documentclass[fleqn,usenatbib]{mnras}

\usepackage{newtxtext,newtxmath}
\usepackage[T1]{fontenc}
\usepackage{graphicx}
\usepackage{amsmath}
\usepackage{xcolor}
\usepackage{xspace}
\usepackage[normalem]{ulem}

\newcommand{\GlitchPy}{{\tt GlitchPy}\xspace}
\newcommand{\BASTA}{{\tt BASTA}\xspace}
\newcommand{\CoRoT}{{\it CoRoT}\xspace}
\newcommand{\Ktwo}{{\it K2}\xspace}
\newcommand{\Kepler}{{\it Kepler}\xspace}
\newcommand{\TESS}{{\it TESS}\xspace}

\def\Teff{T_{\rm eff}}
\def\FeH{[{\rm Fe}/{\rm H}]}
\def\Dnu{\Delta\nu}
\def\amphe{\langle A_{\rm He} \rangle}
\def\delhe{\Delta_{\rm He}}
\def\tauhe{\tau_{\rm He}}
\def\taucz{\tau_{\rm CZ}}
\def\gr{gr_{012}}

\title[Advanced asteroseismic modelling]{Advanced asteroseismic modelling: breaking the degeneracy between stellar mass and initial helium abundance}

\author[Verma et al.]{Kuldeep Verma,$^{1,2,3}$\thanks{E-mail: verma@ice.csic.es}
Jakob L. R\o rsted,$^{2}$\thanks{Formerly Jakob R. Mosumgaard}
Aldo M. Serenelli,$^{1,3}$
Víctor Aguirre B\o rsen-Koch,$^{2}$\thanks{Formerly Víctor Silva Aguirre}
\newauthor
Mark L. Winther,$^{2}$
Amalie Stokholm$^{4,2,5}$
\\
$^1$Instituto de Ciencias del Espacio (ICE, CSIC), Campus UAB, Carrer de Can Magrans, s/n, 08193 Cerdanyola del Valles, Spain\\
$^2$Stellar Astrophysics Centre, Department of Physics and Astronomy, Aarhus University, Ny Munkegade 120, DK-8000 Aarhus C, Denmark\\
$^3$Institut d'Estudis Espacials de Catalunya (IEEC), Gran Capita 4, E-08034, Barcelona, Spain\\
$^4$Dipartimento di Fisica e Astronomia, Universit\`{a} degli Studi di Bologna, Via Gobetti 93/2, I-40129 Bologna, Italy\\
$^5$INAF -- Osservatorio di Astrofisica e Scienza dello Spazio di Bologna, Via Gobetti 93/3, I-40129 Bologna, Italy
}

\date{Accepted XXX. Received YYY; in original form ZZZ}
\pubyear{2022}

\begin{document}
\label{firstpage}
\pagerange{\pageref{firstpage}--\pageref{lastpage}}
\maketitle

% Abstract of the paper
\begin{abstract}
Current stellar model predictions of adiabatic oscillation frequencies differ significantly from the corresponding observed frequencies due to the non-adiabatic and poorly understood near-surface layers of stars. However, certain combinations of frequencies -- known as frequency ratios -- are largely unaffected by the uncertain physical processes as they are mostly sensitive to the stellar core. Furthermore, the seismic signature of helium ionization provides envelope properties while being almost independent of the outermost layers. We have developed an advanced stellar modelling approach in which we complement frequency ratios with parameters of the helium ionization zone while taking into account all possible correlations to put the most stringent constraints on the stellar internal structure. We have tested the method using the \Kepler benchmark star 16~Cyg~A and have investigated the potential of the helium glitch parameters to constrain the basic stellar properties in detail. It has been explicitly shown that the initial helium abundance and mixing-length parameters are well constrained within our framework, reducing systematic uncertainties on stellar mass and age arising for instance from the well-known anti-correlation between the mass and initial helium abundance. The modelling of six additional \Kepler stars including 16~Cyg~B reinforces the above findings and also confirms that our approach is mostly independent from model uncertainties associated with the near-surface layers. Our method is relatively computationally expensive, however, it provides stellar masses, radii and ages precisely in an automated manner, paving the way for analysing numerous stars observed in the future during the ESA PLATO mission.
\end{abstract}

% Keywords.
\begin{keywords}
asteroseismology -- stars: abundances -- stars: fundamental parameters -- stars: interiors -- stars: solar-type
\end{keywords}

\section{Introduction}
%%%%%%%%%%%%%%%%%%%%%%
\label{sec:intro}
Pressure mode oscillations of the Sun and other solar-type stars contain rich information about their internal structure \citep[see e.g.][]{chap13}. The detection and precise measurements of the oscillation frequencies of stars by the French-led \CoRoT satellite \citep{bagl09}, NASA's \Kepler/\Ktwo telescope \citep{boru09,howe14} and more recently by the \TESS mission \citep{rick14} allow the study of stellar interiors in great detail and help in determining the fundamental stellar properties including mass, radius and age to unprecedented precision. Unfortunately, the physical description of the near-surface layers of stars has shortcomings and therefore one-dimensional models of stellar evolution predict frequencies which show frequency-dependent offset from the observed frequencies \citep[the so-called "surface effect"; see e.g.][]{jcd91a}. For this reason, the stellar parameter inference using asteroseismology becomes a challenging problem, requiring careful treatment of the systematic uncertainties arising due to the surface effect.

Stellar modellers use different methods to deal with the surface effect. One of the popular approaches is to correct the model frequencies (semi-)empirically \citep[for various proposed corrections, see][]{kjel08,ball14,sono15}, and compare the corrected model frequencies with the observed ones. Despite such corrections, it turns out that several model frequencies are still off by more than $5\sigma$ due to the high precision of the seismic data. Consequently, the seismic contribution to the $\chi^2$-statistic is much larger than the total number of the observed modes. The optimization in such a case may lead to an incorrect solution as it has the potential to over-fit the residual systematic uncertainty associated with the surface-effect corrected model frequencies at the cost of under-fitting the non-seismic data such as effective temperature, $\Teff$, and surface metallicity, $\FeH$. To moderate the effect of the residual systematic uncertainty, the weight of the seismic term in the definition of the $\chi^2$-statistic is artificially reduced (often by dividing it with the number of observed modes) relative to the other terms \citep[see e.g.][]{cunh21}. Such ad-hoc weights in the definition of the cost function are not statistically justified and can have important implications; for instance, the inferences are no longer guaranteed to be maximum likelihood estimates even if the statistical uncertainties on observables are Gaussian distributed. Furthermore, this approach faces other well-known issues such as giving biased estimates of the initial helium abundance, $Y_i$; for example, \citet{math12} used such a method to study 22 \Kepler stars and found $Y_i$ for 6 stars significantly below the standard Big Bang nucleosynthesis value \citep[$0.2471\pm0.0003$;][]{plan20}. Recently, \citet{jorg18} developed an approach in which they replaced the outermost layers of one-dimensional stellar models by horizontally averaged structure from more realistic three-dimensional hydrodynamic simulations \citep[see also][]{mosu20,jorg21}. However, this is not adequate either because this ignores non-adiabatic effects in the frequency calculations.

There are certain combinations of frequencies, referred to as frequency ratios or simply ratios, which have been shown to be sensitive to the properties of the stellar core while being almost independent of the conditions in the near-surface layers \citep[see e.g.][]{roxb03,oti05,silv11}. The insensitivity of ratios against changes in properties of the near-surface layers makes model ratios free from the systematic uncertainties related to the surface effects. Therefore, model ratios can be directly compared with the corresponding observed values. The frequency ratios have been successfully used to constrain properties of the innermost layers, particularly the amount of central mixed mass for the stars with convective cores \citep[see e.g.][]{silv13,dehe16}. Although using ratios instead of frequencies seems like a promising strategy, the disadvantage is that they are not just insensitive to the outermost layers but also to a large fraction of the whole envelope; most of which do not contribute to the surface effect. Furthermore, the derived ratios have larger fractional uncertainties compared to the observed frequencies. As a result, the methods based on ratios as the only seismic observables have less constraining power compared to those based directly on the individual mode frequencies. For this reason, mode frequencies and the ratios derived from them are sometimes na\"ively fitted together without even taking the correlations among them into account. Such approaches alleviate the sub-primordial $Y_i$ issue to some extent \citep[see e.g.][]{silv13,metc14}, though it is known to be still biased towards lower values \citep[see e.g.][]{metc15}.      

There are certain regions inside solar-type stars where the sound speed or its derivatives change over length scales shorter than the local wavelengths of the pressure mode oscillations; in other words, there are glitches in the acoustic structure of such stars. These glitches leave their signatures on the stellar oscillation frequencies \citep[see e.g.][]{goug88a,voro88,goug90a}. There are two well-studied acoustic glitches; one appears to arise from the localized peak in the first adiabatic index, $\Gamma_1$, between the two stages of helium ionization \citep[He glitch; see][]{broo14,verm14b}, whereas the other emerges due to the near-discontinuity in the gradient of the density scale height at the base of the envelope convection zone \citep[CZ glitch; see e.g.][]{houd07}. Since the average amplitude of the CZ glitch signature ($\sim 0.1\mu$Hz) is typically comparable to the measurement uncertainties on the observed mode frequencies, its analysis suffers from the so-called aliasing problem \citep{mazu01}, making it difficult to reliably determine the properties of the base of the envelope convection zone \citep[see e.g.][]{mazu14,verm17}. In any case, the CZ glitch lies deep enough to leave its signature in the frequency ratios \citep[see e.g.][]{roxb09}. On the other hand, the amplitude of the He glitch signature is generally larger than the observational uncertainties, enabling reliable measurement of the properties of the helium ionization zone. Since helium ionization occurs at a depth of about 2--3\% of the radius in solar-type stars, the corresponding glitch parameters depend on outer layers and, at the same time, are "nearly" independent of the surface effect \citep[see the discussion in Section~\ref{sec:surface_effect} and appendix B of][]{verm19a}. In other words, the He glitch parameters carry information complementary to those contained in frequency ratios; in particular, the amplitude of the He glitch signature can be used to constrain the envelope helium abundance, $Y_s$ \citep[see e.g.][]{basu04,verm14a,verm19a,farn19,farn20,houd22}.  

The potential of using glitch properties in stellar modelling has not been explored in detail. In a preliminary study, \citet{verm17} fitted the He glitch parameters together with just a few frequency ratios (to avoid correlations) and spectroscopic observables ($\Teff$ and $\FeH$). \citet{farn19} developed a method based on Gram-Schmidt's orthogonalisation to post-process the seismic data and extract uncorrelated indicators for ratios and glitch amplitudes, and used it to study 16 Cyg A \& B \citep{farn20}. Both of the above approaches try to avoid correlated observables at the cost of losing information contained in the original data. Just to give an example, \citet{farn19} lose information about the width of the helium ionization zone in the process of linearization. In the present study, we develop a method in which all the observed ratios and He glitch parameters are fitted together with spectroscopic observables. The method takes into account all possible correlations among the ratios and the He glitch parameters. We test our modelling approach on the \Kepler benchmark star 16 Cyg A and discuss its novel aspects in detail. The method is further used to study a sample of \Kepler stars, which clearly demonstrates that, within this framework, all the basic stellar parameters including the initial helium abundance and mixing-length parameter are well constrained, thereby reducing systematic uncertainties on the inferred stellar mass and age.

\section{The data}
%%%%%%%%%%%%%%%%%%
\label{sec:data}
We used the observed effective temperature, surface metallicity and the oscillation frequencies from the \Kepler asteroseismic LEGACY project \citep{lund17a,silv17}. \citet{roxb17} noted a few anomalies in the original seismic data from the LEGACY project, which were later corrected by \citet{lund17b}. In this study, we used the revised set of oscillation frequencies from the LEGACY project. 

Since we need to estimate the combined covariance matrix for the frequency ratios and He glitch parameters, we will calculate these observables along with the corresponding covariance matrix consistently using Monte Carlo (MC) simulations of the observed frequencies (see Section~\ref{sec:covariance}). The lowest and highest observed frequencies are determined using low signal-to-noise data and therefore have associated not only large statistical uncertainty but also large unknown systematic uncertainty. Moreover, statistical uncertainties on such frequencies may not follow a Gaussian distribution as assumed in Sec.~\ref{sec:covariance} while performing the MC simulations to determine the covariance matrix. Therefore, we will not consider frequencies at the extreme ends with observational uncertainties greater than $1.5\mu$Hz in this study.

16 Cyg A \& B are among the \Kepler benchmark solar analog stars with the highest quality asteroseismic data. They have masses in the ranges 1.05--1.11 M$_\odot$ and 0.99--1.03 M$_\odot$, respectively \citep[see e.g.][]{silv17}. We use the observed data for 16 Cyg A \& B to test our modelling approach. In addition, we study KIC 3427720, 6106415, 8379927, 9139151 and 10644253 using our method (see Section~\ref{sec:sample} for the justification of their choice). The observables used in the modelling for all the stars are listed in Table~\ref{tab:tab1}.

\section{The stellar modelling approach}
%%%%%%%%%%%%%%%%%%%%%%%%%%%%%%%%%%%%%%%%
\label{sec:approach}
We adopt a grid-based modelling approach which requires a pre-computed grid of stellar models. The GARching STellar Evolution Code \citep[GARSTEC;][]{weis08} was employed to compute a dense grid of models. The code was used with OPAL atomic opacities \citep{igle96} supplemented by low-temperature opacities of \citet{ferg05}. We exploited the solar metallicity mixture from \citet{aspl09}. The OPAL equation of state was used \citep{roge02}. All the reaction rates were from NACRE except $^{14}{\rm N}(p,\gamma)^{15}{\rm O}$ for which the rate from \citet{form04} were used. We included atomic diffusion and convective core overshoot following the prescriptions of \citet{thou94} and \citet{frey96}, respectively. The adiabatic oscillation frequencies for models in the grid were calculated using the Aarhus adiabatic oscillation package \citep[ADIPLS;][]{jcd08a}.

We wanted to test our fitting method on 16 Cyg A \& B, and generated a dense grid containing about 9,000 tracks over a parameter space suitably chosen for these two stars. Specifically, the grid spans the following ranges: mass $M \in [0.9, 1.2]$ M$_\odot$, initial helium abundance $Y_i \in [0.23, 0.33]$, initial metallicity $[{\rm Fe}/{\rm H}]_i \in [-0.1, 0.3]$ dex, mixing-length $\alpha_{\rm MLT} \in [1.5, 2.1]$ and exponential core overshoot $f_{\rm OV} \in [0.00, 0.03]$. This space was populated uniformly using a quasi-random number generator \citep{sobo67}. Randomly populated grids sample the space more efficiently than evenly spaced grids for problems in which all the parameters defining the space are not equally important (the performance of the two approaches are similar for the rare problems in which all the parameters play equally important role). This can be easily understood if we consider two grids -- one randomly populated and other evenly spaced -- defined on the same set of parameters ($P_1$, $P_2$, ..., $P_n$) and containing the same number of grid points. If it turns out that one parameter, say $P_1$, is irrelevant for the problem at hand, the points in the evenly spaced grid with the same values of $P_2$, $P_3$, ..., $P_n$ but different values of $P_1$ are effectively identical. This clearly does not happen for the randomly populated grid which leads to a denser grid in the relevant parameter space (i.e. the space formed by $P_2$, $P_3$, ..., $P_n$). We computed a few hundred models for each track in the large frequency separation \citep[calculated using the scaling relations,][]{kjel95} range, $[90, 130]$ $\mu$Hz. Note that we chose the large separation to conveniently bracket the models of 16 Cyg A \& B, however one could have used the central hydrogen abundance for this purpose as well. In total, the above resulted in about 2.3 million models. Note that 16 Cyg A \& B are not expected to have convective core, however, we included overshoot as a variable parameter in the grid anticipating that some models with masses at the higher end may have convective core. Furthermore, certain models with masses and compositions close to that of 16 Cyg A may preserve the pre-main sequence convective core for sufficiently large values of the overshoot parameter. It turns out that the other stars studied in Section~\ref{sec:sample} also do not have convective core, and hence models with convective core are mostly redundant in the current work and overshoot acts like a nuisance parameter (or $P_1$ as discussed above). However, this will be an interesting parameter in future asteroseismic studies using our method involving relatively massive stars which have convective core.  

We use the publicly-available software, the BAyesian STellar Algorithm\footnote{\url{https://github.com/BASTAcode/BASTA}} \citep[\BASTA;][]{silv22}, for the stellar parameter inference problem. \BASTA is a versatile code and has the capability to fit a variety of input data including those obtained through spectroscopy, photometry, astrometry, and asteroseismology, allowing precise inferences of stellar properties such as mass, radius and age. We refer the reader to the above-mentioned paper and corresponding online documentation for a more technical description of the Bayesian framework as well as for lists of all possible input observables and output stellar properties. In the following, we will briefly define the seismic observables fitted in this study including the frequency ratios and the He glitch parameters, and write the expression for the likelihood function. 

For a spherically symmetric star, a mode of solar-like oscillation is characterized by the harmonic degree, $l$, and radial order, $n$. We denote the corresponding mode frequency by $\nu_{n,l}$. For a typical solar-like oscillator with the observed radial, dipole and quadrupole mode frequencies, we can calculate the three most popular ratios:
\begin{equation}
r_{02} (n) = \frac{\nu_{n,0} - \nu_{n-1,2}}{\nu_{n,1} - \nu_{n-1,1}},
\label{eq:r02}
\end{equation}
\begin{equation}
r_{01} (n) = \frac{\nu_{n-1,0} - 4\nu_{n-1,1} + 6\nu_{n,0} - 4\nu_{n,1} + \nu_{n+1,0}}{8 (\nu_{n,1} - \nu_{n-1,1})},
\label{eq:r01}
\end{equation}
and
\begin{equation}
r_{10} (n) = - \frac{\nu_{n-1,1} - 4\nu_{n,0} + 6\nu_{n,1} - 4\nu_{n+1,0} + \nu_{n+1,1}}{8 (\nu_{n+1,0} - \nu_{n,0})}. 
\label{eq:r10}
\end{equation}
\citet{roxb18} pointed out that ratios $r_{01}$ and $r_{10}$ contain the same information about the stellar interior and hence should not be used together in the modelling. He suggested to either combine $r_{01}$ and $r_{02}$ into the sequence $r_{012} = \{r_{01}(n), r_{02}(n), r_{01}(n+1), r_{02}(n+1), \dots\}$ or equivalently $r_{10}$ and $r_{02}$ into $r_{102} = \{r_{02}(n), r_{10}(n), r_{02}(n+1), r_{10}(n+1), \dots\}$. In \BASTA, the user can choose to fit any of the above ratios, however in this study, we have decided to fit $r_{012}$ following the suggestion of \citet{roxb18}. Furthermore, following \citet{roxb13} we interpolate the model ratios at the corresponding observed frequencies before comparing them with the observed ratios.

\begin{figure*}
	\includegraphics[width=\textwidth]{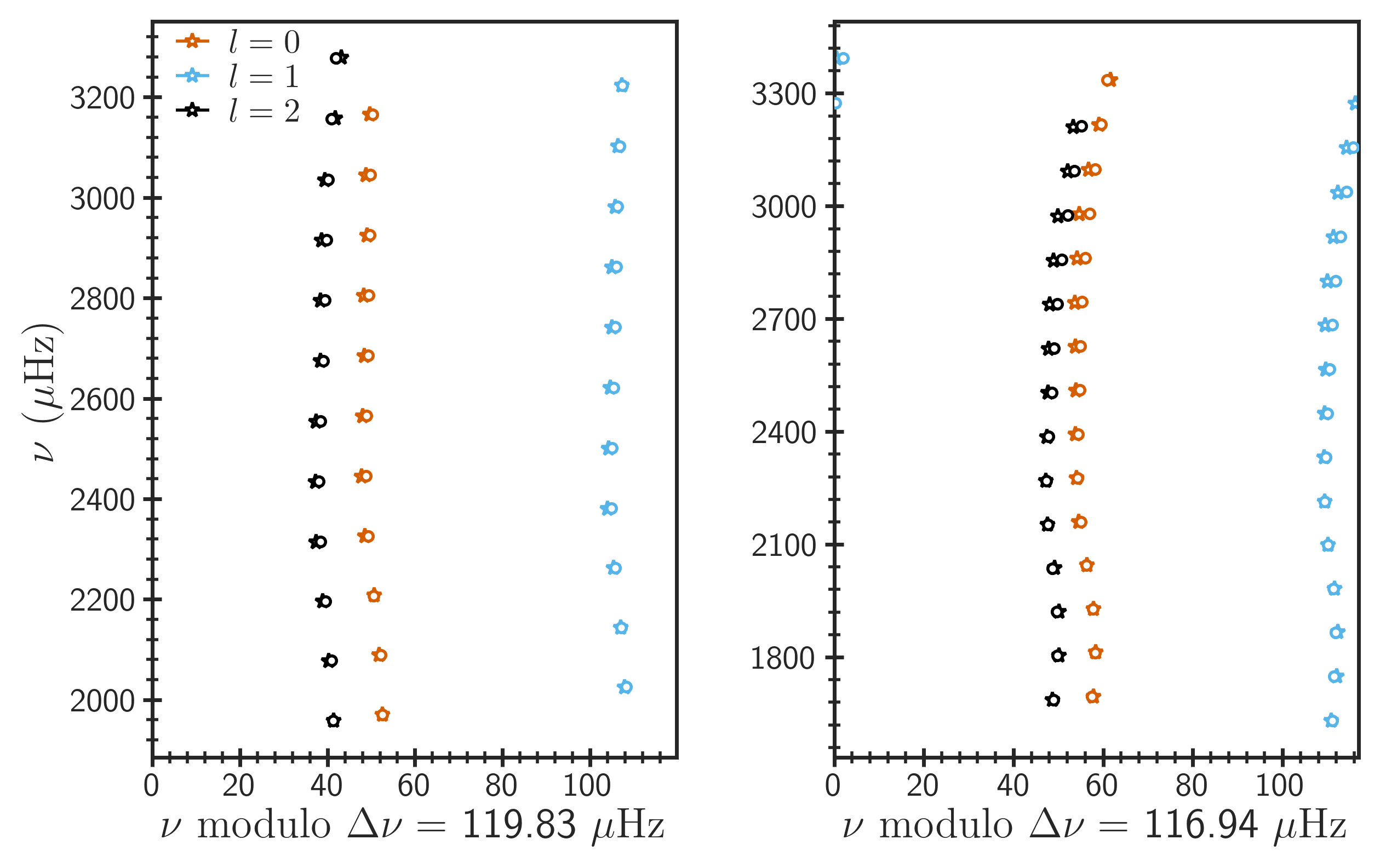}
    \caption{\'Echelle diagrams for KIC~3427720 (left panel) and 16~Cyg~B (right panel). The star symbols represent the observed data while the circles show the corresponding best-fitting models. The model frequencies were corrected for the surface effect. The colours indicate different harmonic degrees as shown in the legend.}
    \label{fig:fig1}
\end{figure*}

The total contribution of the He and CZ glitches to the oscillation frequencies can be written as \citep[see e.g.][]{houd07}: 
\begin{eqnarray}
\delta\nu &=& A_{\rm He} \nu e^{-8\pi^2\delhe^2\nu^2} \sin(4\pi\tauhe\nu + \psi_{\rm He})\nonumber\\
&+&\frac{A_{\rm CZ}}{\nu^2} \sin(4\pi\taucz\nu + \psi_{\rm CZ}),
\label{eq:glh_fq}
\end{eqnarray}
where the first term on the right hand side represents contribution from the He glitch while the second term from the CZ glitch. The parameters $A_{\rm He}$ and $A_{\rm CZ}$ measure the amplitudes of the corresponding glitch signatures, $\delhe$ the acoustic width of the peak in $\Gamma_1$ between the two stages of He ionization, $\tauhe$ and $\taucz$ measure the acoustic depths of the respective glitches, and $\psi_{\rm He}$ and $\psi_{\rm CZ}$ are the phase constants. These parameters are determined either by fitting the oscillation frequencies directly or by fitting the second differences of frequencies with respect to the radial order. Note that although Eq.~\ref{eq:glh_fq} was introduced as the contributions of glitches to the frequencies, the functional form remains the same for contributions to the second differences, though amplitudes and phases have slightly different interpretations and hence different values \citep[see e.g.][]{houd07}. The amplitudes in second differences can be converted to the amplitudes in frequencies by dividing them with $4\sin^2(2\pi\tau_g\Dnu)$, where $\tau_g$ and $\Dnu$ are the corresponding acoustic depth of the glitch and the large frequency separation, respectively \citep[see e.g.][]{basu94}. We have presented one particular implementation of fitting each, the frequencies (Method A) and the second differences (Method B), in \citet{verm19a}, and have now made them publicly available with the python code, \GlitchPy\footnote{\url{https://github.com/kuldeepv89/GlitchPy}}. We provide a more complete description of both methods in the appendix (see Sections~\ref{sec:methoda} and \ref{sec:methodb}). 

The Method A was already part of the \BASTA code in the version released with the publication of \citet{silv22}, while the Method B is a new addition. Unless explicitly stated otherwise, we use Method A for determining the glitch properties in this study. In the stellar modelling, we include the He glitch parameters $\delhe$, $\tauhe$ and the so-called average amplitude,
\begin{eqnarray}
\amphe &=& \frac{\int_{\nu_1}^{\nu_2} A_{\rm He} \nu e^{-8\pi^2\delhe^2\nu^2} d\nu}{\int_{\nu_1}^{\nu_2} d\nu}\nonumber\\
&=& \frac{A_{\rm He} [e^{-8\pi^2\delhe^2\nu_1^2} - e^{-8\pi^2\delhe^2\nu_2^2}]}{16\pi^2\delhe^2[\nu_2 - \nu_1]},
\label{eq:amplitude}
\end{eqnarray}
where $\nu_1$ and $\nu_2$ denote the lower and upper frequencies, together defining the frequency range over which to perform the averaging. We choose the minimum and maximum values of the observed frequencies as $\nu_1$ and $\nu_2$ for consistently averaging both the observed and model He glitch amplitudes. We combine the frequency ratios and the He glitch parameters to form the sequence,
\begin{equation}
\gr = \{r_{012}, \amphe, \delhe, \tauhe\}, 
\end{equation}
and include this set of seismic observables as constraints to the inference problem.

Additionally, we include in the fit the large frequency separation, $\Dnu$, calculated using the radial modes following the prescription of \citet{whit11}. Note that this is a quantity which depends on the surface effect to some extent; for example, its value for a solar model is systematically larger than the Sun by about $1\mu$Hz \citep[see e.g.][]{kjel08}. Therefore, before computing $\Dnu$, we correct the model frequencies for the surface effect using the power-law correction of \citet{kjel08}, and subsequently use it in the modelling with caution (see below). We do not use the currently more popular correction of \citet{ball14} because it does not limit its correction parameters and thus can potentially (over-)correct for large systematic differences between the model and observation throughout the frequency spectrum (including the lowest frequencies for which the surface effect contributes a maximum of only a few microhertz in the case of the Sun). Since the value of the exponent, $b$, in \citet{kjel08} correction depends on the number of observed modes or more specifically on the frequency range in units of the frequency of maximum power \citep{sono15}, we calculated this range for all the stars in our sample. It turns out that the lower and upper limits of this range vary little with the star ($0.65\pm0.03$ -- $1.29\pm0.08$), and suggest a lower value of $b$ \citep[slightly less than 3 according to Figure~14 of][]{sono15} than what is typically assumed \citep[4.40--5.25; see e.g.][]{kjel08}. In our study, we assumed $b = 3$ for all the stars. Clearly $b$ and hence model $\Dnu$ have associated systematic uncertainty which should be taken into account while modelling an individual star. We assumed an uncertainty of $0.2\mu$Hz on model $\Dnu$ while defining the likelihood function (see the next paragraph). In Figure~\ref{fig:fig1}, we show the so-called \'Echelle diagrams for two stars; one shows the best case scenario (KIC 3427720) for which the observed and surface-corrected best-fitting model $\Dnu$ values differ only by $0.03\mu$Hz, whereas the other demonstrates the worse case (16~Cyg~B) for which they differ by $0.24\mu$Hz (for the best-fitting models, see Section~\ref{sec:sample}). 

To fit the above set of seismic observables together with the spectroscopic data, we define a likelihood function $P$ as,
\begin{equation}
P({\bmath D} | {\bmath \Theta}) \propto \exp\left(-\chi^2 / 2\right),
\label{eq:likelihood}
\end{equation}
where ${\bmath D}$ and ${\bmath \Theta}$ represent the data and model parameters, respectively. The $\chi^2$ is defined as,
\begin{eqnarray}
\chi^2 &=& \left(\frac{T_{\rm eff,o} - T_{\rm eff,m}}{\sigma_{\Teff}}\right)^2 + \left(\frac{[{\rm Fe}/{\rm H}]_{\rm o} - [{\rm Fe}/{\rm H}]_{\rm m}}{\sigma_{\FeH}}\right)^2\nonumber\\
&+& \left(\frac{\Delta\nu_{\rm o} - \Delta\nu_{\rm m}}{\sqrt{\sigma_{\Dnu}^2 + \sigma_{\Dnu,{\rm m}}^2}}\right)^2\nonumber\\
&+& \left(\bmath{gr}_{012,{\rm o}} - \bmath{gr}_{012,{\rm m}}\right)^{\rm T} \mathbfss{C}^{-1} \left(\bmath{gr}_{{012,\rm o}} - \bmath{gr}_{012,{\rm m}}\right),
\label{eq:chi2}
\end{eqnarray}
where $\sigma_{\Teff}$, $\sigma_{\FeH}$ and $\sigma_{\Dnu}$ are observational uncertainties on $\Teff$, $\FeH$ and $\Dnu$, respectively, while $\mathbfss{C}$ is the covariance matrix for $\gr$. We assume $\sigma_{\Dnu,{\rm m}} = 0.2\mu$Hz as discussed in the previous paragraph. We do not consider $\Dnu$ as part of $\gr$ because of its special treatment for the associated systematic uncertainty (further justification is given in Section~\ref{sec:covariance}). As presented in Sections~\ref{sec:methoda} and \ref{sec:methodb}, the extraction of glitch parameters from the oscillation frequencies or second differences involves a non-linear optimization in a high-dimensional space, making the calculation of the likelihood a computationally expensive task. On average, it takes about 0.3 second on a modern desktop to calculate the likelihood of one model, which means it would take about one week to model a single star if we evaluate all 2.3 million models in the grid. To improve the computational efficiency, we only compute the likelihood for those models for which $\Teff$ and $\FeH$ agree with the corresponding observed values within $5\sigma$ and 0.25~dex and the frequency of the observed radial mode with the smallest $n$ value and corresponding model mode frequency are in agreement within 15\% of the observed $\Dnu$. This reduces the number of models for which the likelihood need to be evaluated to 60,000--100,000, depending on the star. We combine the likelihood together with an initial mass function \citep{salp55} as a prior on mass to compute the posterior probability and subsequently derive the stellar parameters and associated uncertainties following \citet{silv22}. 

\begin{figure}
	\includegraphics[width=\columnwidth]{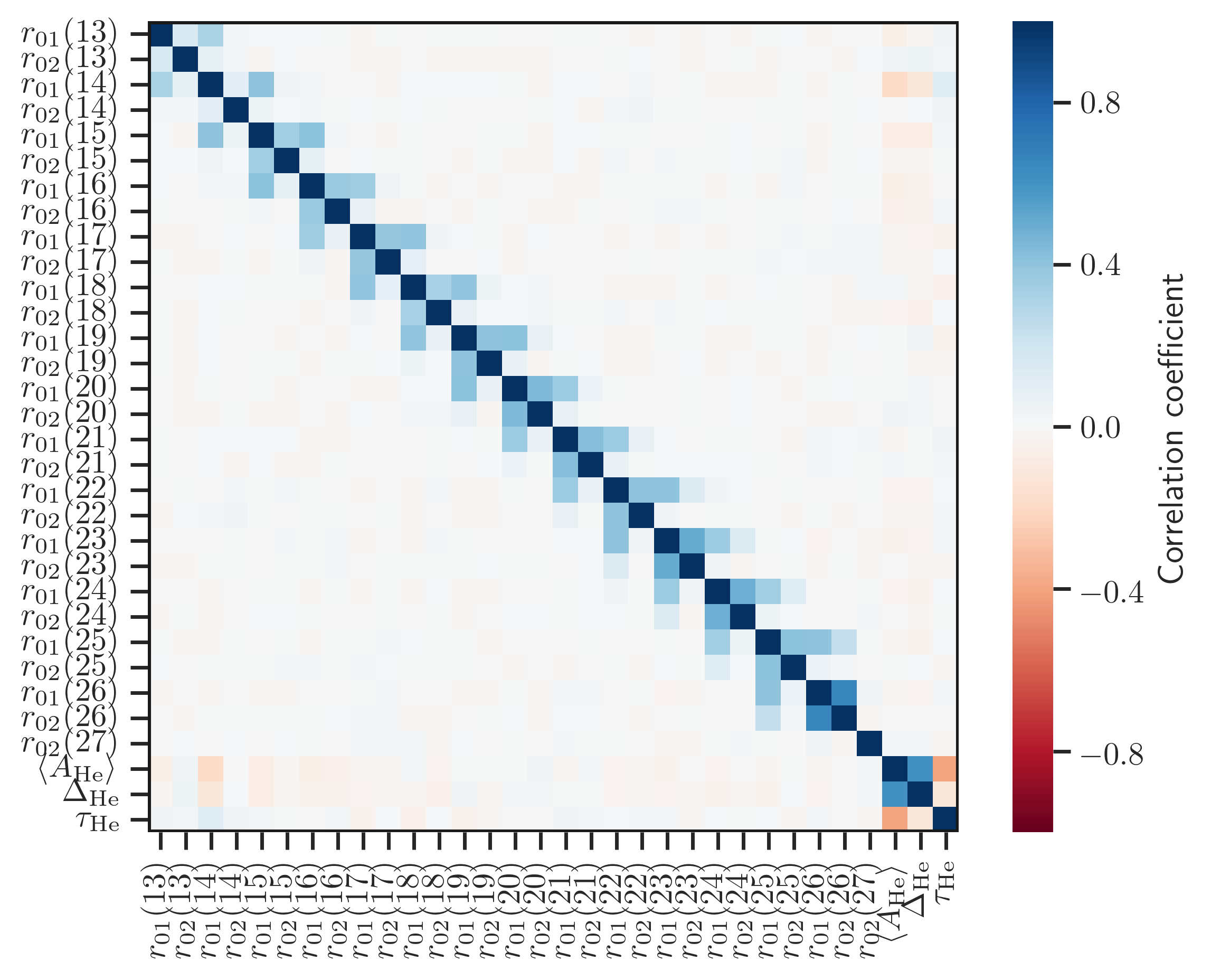}
    \caption{Correlation matrix for the frequency ratios and He glitch parameters for 16 Cyg A. The colors represent the values of the Pearson correlation coefficient. We plot the correlation matrix instead of covariance matrix for clarity.}
    \label{fig:fig2}
\end{figure}

\section{Results}
%%%%%%%%%%%%%%%%%
\label{sec:results}
We now test the above stellar modelling approach using the data for 16 Cyg A, and also demonstrate how well it constrains various stellar properties including the initial helium abundance and mixing-length for not just 16 Cyg A but also for a sample of other \Kepler stars. However, we first need to estimate the covariance matrix for $\gr$ followed by its inverse, which will allow us to compute $\chi^2$ as defined in Eq.~\ref{eq:chi2} and subsequently the likelihood in Eq.~\ref{eq:likelihood}.

\subsection{Covariance matrix and its inverse}
%%%%%%%%%%%%%%%%%%%%%%%%%%%%%%%%%%%%%%%%%%%%%%
\label{sec:covariance}
Since the same mode frequency is used when computing multiple ratios (see Eqs. \ref{eq:r02}--\ref{eq:r10}), the ratios are expected to be correlated. Furthermore, since the He glitch properties are also derived from the same set of frequencies, in addition to the internal correlations among the glitch parameters, there may be possible correlations between the ratios and the He glitch parameters (as we will see later in this section, they turn out to be small). It is important to take into account these correlations when making the parameter inferences, and therefore we include them by using the covariance matrix in Eq. \ref{eq:chi2}.

\begin{figure}
	\includegraphics[width=\columnwidth]{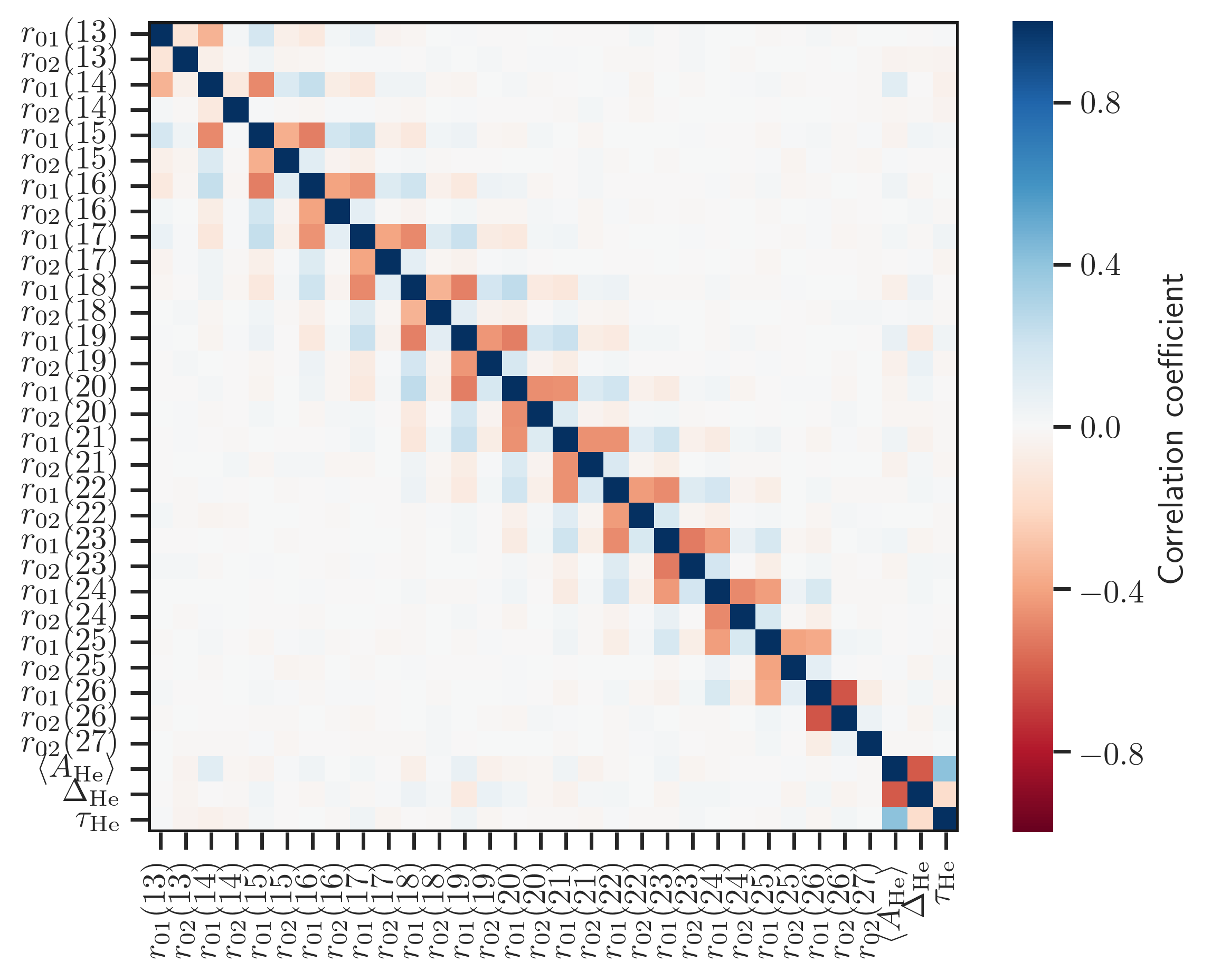}
    \caption{The same as Figure~\ref{fig:fig2} but for the inverse of the covariance matrix for 16 Cyg A.}
    \label{fig:fig3}
\end{figure}
\begin{table*}
	\centering
	\caption{All the observables used in the stellar model fitting for different stars (columns 2--8). The observed effective temperature and metallicity are from the LEGACY project \citep[see][and references therein]{lund17a,silv17}. The uncertainties on the frequency ratios and He glitch parameters are from the computed covariance matrices.}
	\label{tab:tab1}
	\scalebox{0.85}{
	\begin{tabular}{cccccccc}
		\hline
		Observable       & KIC 3427720      & KIC 6106415      & KIC 8379927      & KIC 9139151     & KIC 10644253    & 16 Cyg A        & 16 Cyg B\\
		\hline
		$\Teff$ (K)      &   $6045\pm77$    &   $6037\pm77$    &   $6067\pm120$   &  $6302\pm77$    &  $6045\pm77$    &  $5825\pm50$    &  $5750\pm50$\\
		$\FeH$ (dex)     & $-0.060\pm0.100$ & $-0.040\pm0.100$ & $-0.100\pm0.150$ & $0.100\pm0.100$ & $0.060\pm0.100$ & $0.100\pm0.026$ & $0.050\pm0.021$\\
		$\Dnu$ ($\mu$Hz) & $119.83\pm0.12$  & $104.07\pm0.11$  & $120.27\pm0.11$  & $117.15\pm0.12$ & $122.78\pm0.11$ & $103.28\pm0.10$ & $116.94\pm0.12$\\
		$r_{01} (13)$ &         \dots       &         \dots       &         \dots       &         \dots       &        \dots        & $0.05028\pm0.00283$ &         \dots      \\
		$r_{02} (13)$ &         \dots       &         \dots       &         \dots       &         \dots       &        \dots        & $0.06892\pm0.00705$ & $0.07811\pm0.00569$\\
		$r_{01} (14)$ &         \dots       & $0.04284\pm0.00156$ &         \dots       &         \dots       &        \dots        & $0.04772\pm0.00103$ & $0.03802\pm0.00122$\\
		$r_{02} (14)$ &         \dots       & $0.07611\pm0.00327$ &         \dots       &         \dots       &        \dots        & $0.08078\pm0.00383$ & $0.07001\pm0.00506$\\
		$r_{01} (15)$ &         \dots       & $0.04271\pm0.00116$ & $0.03464\pm0.00234$ &         \dots       &        \dots        & $0.04582\pm0.00100$ & $0.03634\pm0.00090$\\
		$r_{02} (15)$ &         \dots       & $0.07560\pm0.00364$ &         \dots       &         \dots       &        \dots        & $0.06873\pm0.00206$ & $0.06579\pm0.00177$\\
		$r_{01} (16)$ & $0.03538\pm0.00185$ & $0.04005\pm0.00101$ & $0.03584\pm0.00188$ & $0.03848\pm0.00360$ & $0.02931\pm0.00699$ & $0.04266\pm0.00076$ & $0.03403\pm0.00064$\\
		$r_{02} (16)$ & $0.09632\pm0.00570$ & $0.06930\pm0.00209$ &         \dots       &         \dots       &         \dots       & $0.06396\pm0.00149$ & $0.06190\pm0.00163$\\
		$r_{01} (17)$ & $0.03566\pm0.00259$ & $0.03968\pm0.00089$ & $0.03589\pm0.00134$ & $0.03266\pm0.00202$ & $0.03440\pm0.00357$ & $0.03998\pm0.00056$ & $0.02941\pm0.00056$\\
		$r_{02} (17)$ & $0.09899\pm0.00586$ & $0.07185\pm0.00192$ & $0.09531\pm0.00409$ &         \dots       &         \dots       & $0.05944\pm0.00108$ & $0.06011\pm0.00107$\\
		$r_{01} (18)$ & $0.03321\pm0.00138$ & $0.03662\pm0.00082$ & $0.03395\pm0.00098$ & $0.03106\pm0.00182$ & $0.03271\pm0.00289$ & $0.03708\pm0.00047$ & $0.02766\pm0.00048$\\
		$r_{02} (18)$ & $0.09509\pm0.00340$ & $0.07104\pm0.00145$ & $0.09899\pm0.00319$ & $0.10423\pm0.00969$ & $0.10461\pm0.00857$ & $0.05749\pm0.00088$ & $0.05852\pm0.00092$\\
		$r_{01} (19)$ & $0.03105\pm0.00099$ & $0.03490\pm0.00068$ & $0.03138\pm0.00086$ & $0.03383\pm0.00133$ & $0.03194\pm0.00173$ & $0.03484\pm0.00044$ & $0.02577\pm0.00039$\\
		$r_{02} (19)$ & $0.08832\pm0.00158$ & $0.06850\pm0.00115$ & $0.09154\pm0.00188$ & $0.09880\pm0.00551$ & $0.10074\pm0.01031$ & $0.05378\pm0.00066$ & $0.05463\pm0.00069$\\
		$r_{01} (20)$ & $0.02961\pm0.00105$ & $0.03219\pm0.00063$ & $0.03206\pm0.00076$ & $0.03195\pm0.00099$ & $0.03206\pm0.00140$ & $0.03232\pm0.00052$ & $0.02425\pm0.00038$\\
		$r_{02} (20)$ & $0.08852\pm0.00212$ & $0.06826\pm0.00112$ & $0.09234\pm0.00146$ & $0.08628\pm0.00299$ & $0.09711\pm0.00269$ & $0.05123\pm0.00086$ & $0.05344\pm0.00066$\\
		$r_{01} (21)$ & $0.02858\pm0.00078$ & $0.03052\pm0.00077$ & $0.03147\pm0.00081$ & $0.02928\pm0.00089$ & $0.03385\pm0.00128$ & $0.02848\pm0.00057$ & $0.02253\pm0.00042$\\
		$r_{02} (21)$ & $0.08280\pm0.00145$ & $0.06554\pm0.00139$ & $0.08975\pm0.00145$ & $0.08553\pm0.00239$ & $0.09851\pm0.00186$ & $0.04607\pm0.00092$ & $0.05047\pm0.00071$\\
		$r_{01} (22)$ & $0.02767\pm0.00114$ & $0.02952\pm0.00098$ & $0.03127\pm0.00076$ & $0.02808\pm0.00107$ & $0.03385\pm0.00165$ & $0.02641\pm0.00081$ & $0.02093\pm0.00055$\\
		$r_{02} (22)$ & $0.08287\pm0.00191$ & $0.06427\pm0.00215$ & $0.09276\pm0.00148$ & $0.08206\pm0.00248$ & $0.09321\pm0.00271$ & $0.04507\pm0.00140$ & $0.04786\pm0.00081$\\
		$r_{01} (23)$ & $0.02864\pm0.00142$ & $0.02651\pm0.00134$ & $0.02838\pm0.00088$ & $0.03107\pm0.00140$ & $0.03349\pm0.00152$ & $0.02395\pm0.00137$ & $0.01647\pm0.00086$\\
		$r_{02} (23)$ & $0.08628\pm0.00229$ & $0.05383\pm0.00248$ & $0.08888\pm0.00166$ & $0.08893\pm0.00299$ & $0.09154\pm0.00312$ & $0.04039\pm0.00238$ & $0.04376\pm0.00132$\\
		$r_{01} (24)$ & $0.02303\pm0.00228$ & $0.02611\pm0.00178$ & $0.02656\pm0.00117$ & $0.02982\pm0.00202$ & $0.02778\pm0.00271$ & $0.01726\pm0.00220$ & $0.01287\pm0.00147$\\
		$r_{02} (24)$ & $0.07812\pm0.00352$ & $0.05251\pm0.00376$ & $0.08793\pm0.00252$ & $0.08044\pm0.00467$ & $0.08503\pm0.00394$ & $0.04090\pm0.00418$ & $0.04054\pm0.00294$\\
		$r_{01} (25)$ &         \dots       & $0.02335\pm0.00348$ & $0.02490\pm0.00191$ & $0.02352\pm0.00372$ & $0.02630\pm0.00532$ & $0.01952\pm0.00375$ & $0.01639\pm0.00299$\\
		$r_{02} (25)$ & $0.06514\pm0.00805$ & $0.05291\pm0.00629$ & $0.08622\pm0.00366$ & $0.07949\pm0.01156$ & $0.08574\pm0.00803$ & $0.03416\pm0.00821$ & $0.03930\pm0.00588$\\
		$r_{01} (26)$ &         \dots       & $0.01437\pm0.00636$ & $0.02601\pm0.00244$ &         \dots       &         \dots       & $0.02750\pm0.00763$ & $0.01967\pm0.00397$\\
		$r_{02} (26)$ &         \dots       & $0.03465\pm0.01213$ & $0.09034\pm0.00538$ & $0.08452\pm0.01532$ &         \dots       & $0.03813\pm0.01141$ & $0.04859\pm0.01067$\\
		$r_{01} (27)$ &         \dots       &         \dots       & $0.02862\pm0.00376$ &         \dots       &         \dots       &         \dots       &         \dots      \\
		$r_{02} (27)$ &         \dots       & $0.02737\pm0.01365$ & $0.08558\pm0.00680$ &         \dots       &         \dots       & $0.04353\pm0.01277$ &         \dots      \\
		$r_{01} (28)$ &         \dots       &         \dots       & $0.03652\pm0.00570$ &         \dots       &         \dots       &         \dots       &         \dots      \\
		$r_{02} (28)$ &         \dots       &         \dots       & $0.10479\pm0.00986$ &         \dots       &         \dots       &         \dots       &         \dots      \\	
		$r_{01} (29)$ &         \dots       &         \dots       & $0.01595\pm0.00680$ &         \dots       &         \dots       &         \dots       &         \dots      \\
		$\amphe$ ($\mu$Hz) & $0.527\pm0.051$ & $0.582\pm0.025$ & $0.694\pm0.032$ & $0.685\pm0.061$ & $0.940\pm0.138$ & $0.547\pm0.031$ & $0.470\pm0.019$\\
		$\delhe$ (s)       &  $78.5\pm8.0$   &  $97.7\pm3.8$   &  $78.2\pm3.1$   &  $82.6\pm6.8$   &  $75.9\pm9.4$   & $106.2\pm4.1$   &  $85.6\pm3.4$\\
		$\tauhe$ (s)       & $803.6\pm44.2$  & $874.8\pm21.8$  & $703.8\pm20.8$  & $786.6\pm33.2$  & $666.0\pm47.9$  & $907.4\pm21.2$  & $781.6\pm16.8$\\
		\hline
	\end{tabular}
	}
\end{table*}

To estimate the covariance matrix for $\gr$, we generate 10,000 realizations of the observed frequencies assuming their uncertainties are independent and Gaussian distributed with mean zero and standard deviation given by the corresponding observed uncertainties. Subsequently, we compute $\gr$ by using Eqs.~\ref{eq:r02} and \ref{eq:r01} and performing glitch analysis for all the realizations. In this manner, we have 10,000 realizations of $\gr$ which are used to estimate the covariance matrix. It is possible that the glitch analysis fails to converge or converges to a local minimum instead of the global minimum for some realizations (the number of such realizations are generally small for the stars analysed in this study because of the high precision of the corresponding data). This means that $\gr$ realizations include some outliers, which can potentially affect the covariance matrix significantly. To overcome this problem, we use a covariance estimator that is robust against outliers, known as the minimum covariance determinant estimator\footnote{\url{https://scikit-learn.org/stable/modules/generated/sklearn.covariance.MinCovDet.html}} \citep{pedr11}. We adopt the medians of realizations as values for the corresponding observables.

Figure~\ref{fig:fig2} shows the correlation matrix obtained for 16 Cyg A. We plot the correlation matrix instead of the covariance matrix for clarity (elements of the covariance matrix have different scales). We varied the number of realizations of the observed frequencies from 1,000 to 100,000 and monitored the correlation matrix visually and the variances (diagonal elements) numerically. We found that the matrix converges well for all the stars analyzed in this study by the time number of realizations reach 10,000 (increasing realizations from 5,000 to 10,000 leads to changes in standard deviations below a few percent). Note in the figure that it is symmetric as one would expect for a correlation matrix. Furthermore, we computed eigenvalues of the covariance matrix using singular value decomposition to confirm that it is positive-semidefinite. We also like to note in the figure that the correlations between the ratios and the He glitch parameters are small (typically less than 0.1). However, the trend is noteworthy; the glitch parameters have small but visible correlations with the ratios at the lower frequencies, whereas they are almost independent at the higher frequencies. This is likely due to the fact that the He signature has high significance at the lower frequencies and low significance at the higher frequencies. The results of the above sanity checks for the covariance matrices of other stars studied in this work are the same. 

Table~\ref{tab:tab1} provides values for all the observables included in the definition of $\chi^2$ defined in Eq.~\ref{eq:chi2} with the quoted uncertainties on the frequency ratios and the He glitch parameters from the computed covariance matrices (square root of the corresponding diagonal elements). The ratios and associated uncertainties are in good agreement with the revised data from the LEGACY project \citep{lund17b} as well as those from the \citet{roxb17}. Note that the values of $\amphe$ are slightly different compared to \citet{verm17} for some stars because of the differences in the frequency ranges used in amplitude averaging (see Eq.~\ref{eq:amplitude}). The values of $\tauhe$ are well within $1\sigma$ of the values reported in \citet{verm17} for all the stars (since they did not tabulate $\delhe$, we cannot compare its values). 

\begin{figure*}
	\includegraphics[width=\textwidth]{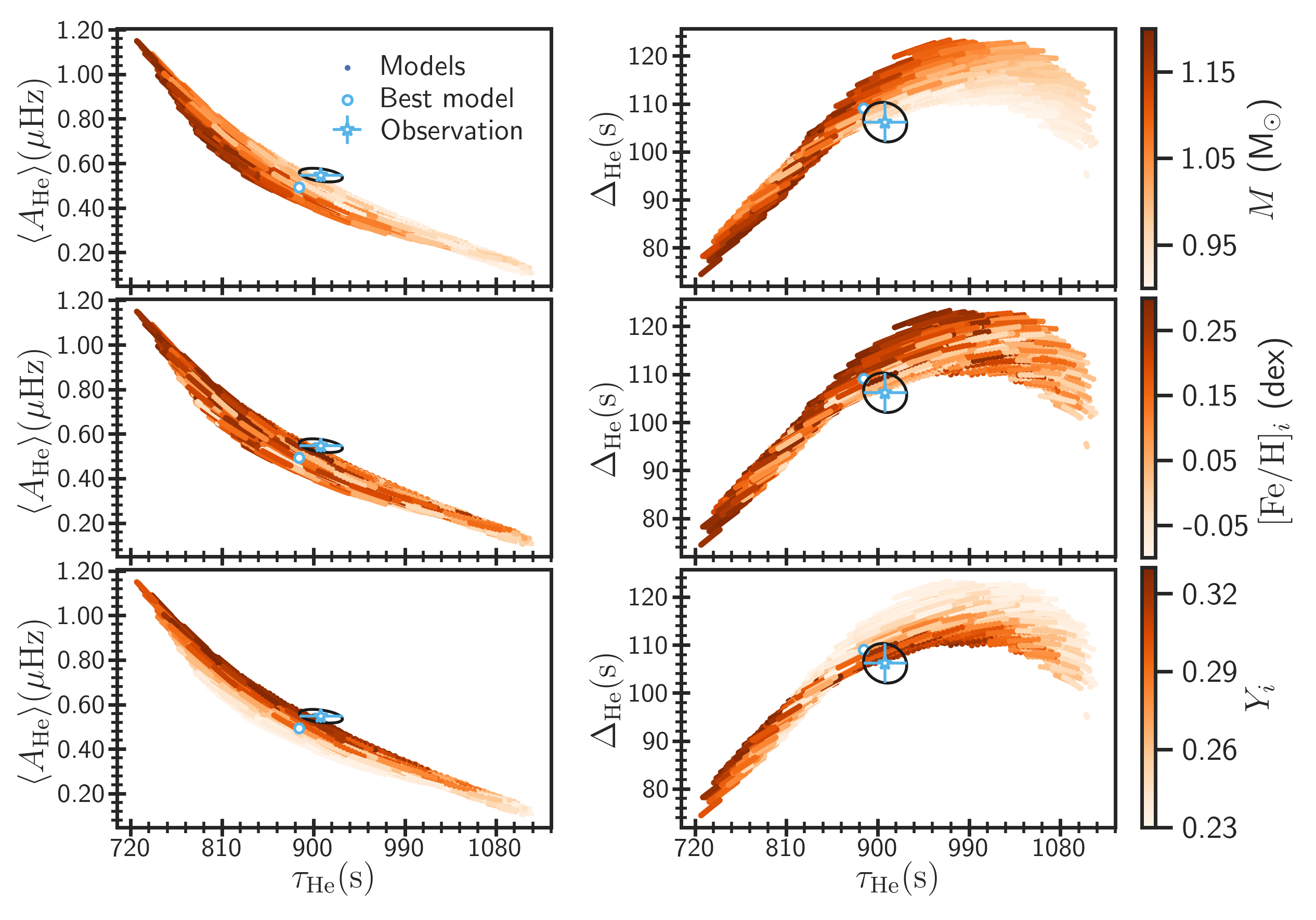}
    \caption{Average amplitude (left panels) and acoustic width (right panels) as a function of acoustic depth for 16 Cyg A. In each panel, the dots represent all the models selected from the grid based on the observed $\Teff$, $\FeH$ and the frequency of the radial mode with the smallest radial order (see the last paragraph of Section~\ref{sec:approach}). Note that a trail represents models from a single track. While performing glitch analysis for the model frequencies, we used the same set of modes and weights in Eq.~\ref{eq:chi2_a} as for the observed data. The open circle highlights the best-fitting model. The star symbol with errorbars represents the observed values while the ellipse enclosing it highlights $1\sigma$ confidence region. The models are colour-coded with the mass (top row), initial metallicity (middle row) and the initial helium abundance (bottom row).}
    \label{fig:fig4}
\end{figure*}

Although the large frequency separation is also derived from the oscillation frequencies, we assume it to be uncorrelated with $\gr$ in Eq.~\ref{eq:chi2}. To test this assumption, we construct the full covariance matrix for 16 Cyg A by including $\Dnu$ as the first element in the vector $\gr$. To account for the surface effect, we added $\sigma_{\Dnu,{\rm m}}^2$ (= 0.04 $\mu$Hz$^2$) in the variance of $\Dnu$ (the corresponding diagonal element in the matrix). The full correlation matrix is shown in Figure~\ref{fig:fig1a}. Clearly, the correlations between $\Dnu$ and other seismic quantities are very small ($< 0.04$). This is mainly because the scale along $\Dnu$ is set by the large systematic uncertainty from the surface effect and any anisotropy remains at smaller scale set by the statistical uncertainty.

The covariance matrix is typically used in parameter inference problems only when the observables are correlated. In such a situation, the matrix is close to being singular, and hence care must be taken when estimating its inverse. While fitting a second-degree polynomial to $r_{010} = \{r_{01}(n), r_{10}(n), r_{01}(n+1), r_{10}(n+1), \dots\}$ to infer the size of the stellar convective core, \citet{dehe16} realized, by observing a systematic difference between the fit and the data, that their estimated inverse covariance matrix was not sufficiently accurate due to ill-conditioning of the original matrix. To remedy the situation, they estimated instead the Moore-Penrose (pseudo-)inverse by setting a few lowest singular values to zero. We wish to point out that this should be avoided, if possible, because suppressing singular values can improve the conditioning of the matrix, however at the cost of changing the original matrix by hand (which may have important implications for the inference). We will also compute pseudo-inverse but suppress only those singular values that are smaller than $10^{-12}$ times the largest singular value. Since condition numbers of covariance matrices of all the stars considered in this study are of the order of $10^{9}$ -- i.e. the ratios of their smallest and largest singular values are of the order of $10^{-9}$ (larger than $10^{-12}$) -- we emphasize that none of the singular values were actually suppressed while calculating the inverses. Figure~\ref{fig:fig3} shows the inverse of the covariance matrix (converted to the corresponding correlation matrix for the plotting purpose) for 16 Cyg A. Although it is difficult to ensure whether this inverse matrix is sufficiently accurate or not, we will see in the following sections that, unlike \citet{dehe16}, we do not encounter any systematic problems while fitting the data.  

\begin{figure*}
	\includegraphics[width=\textwidth]{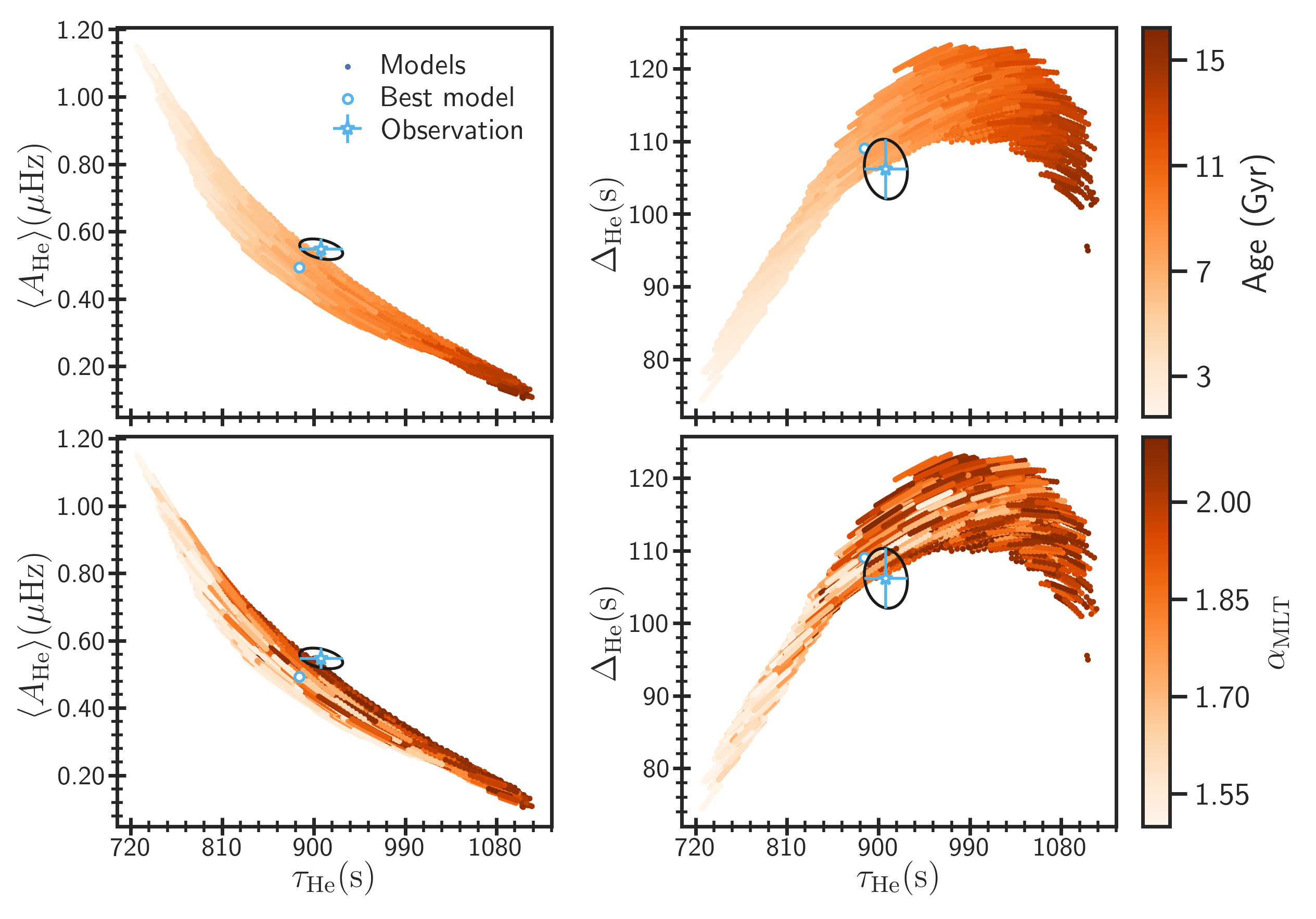}
    \caption{Same as Figure~\ref{fig:fig4}, except now the models are colour-coded with the stellar age (top row) and the mixing-length (bottom row).}
    \label{fig:fig5}
\end{figure*}

\subsection{Test case: 16 Cyg A}
%%%%%%%%%%%%%%%%%%%%%%%%%%%%%%%%
\label{sec:test}
16 Cyg A is the primary component of a binary system with the secondary component being 16 Cyg B. These are among the brightest stars observed by the \Kepler satellite, and are known to exhibit solar-like oscillations. Their highest quality asteroseismic data make them suitable targets for a variety of studies \citep[see e.g.][just to name a few]{metc12,verm14a,buld16a,bell17,farn20}. For 16 Cyg A, the conventional forward modelling approaches based on the spectroscopic and asteroseismic data return mass in the range 1.05--1.11 M$_\odot$, radius in 1.215--1.240 R$_\odot$ and age in the range 6.7--7.5 Gyr, while for 16 Cyg B in ranges 0.99--1.03 M$_\odot$, 1.096--1.111 R$_\odot$ and 6.9--7.4 Gyr, respectively \citep[see e.g.][]{silv17}. \citet{metc15} modelled each component of this binary system independently using the full \Kepler data and found their initial helium abundance to be $0.25\pm0.01$. To determine any biases in their estimates, they carried out a similar analysis of a \Kepler-like data for the Sun, and showed that their $Y_i$ estimates were biased towards lower values by 0.02--0.03. 

We used the approach developed in this work to fit the observables listed in Table~\ref{tab:tab1} for 16~Cyg~A to determine all the free parameters ($M$, $[{\rm Fe}/{\rm H}]_i$, $Y_i$, age $t_{\rm age}$ and $\alpha_{\rm MLT}$). As a sanity check for the proper extraction of the He glitch properties from the model frequencies as well as to closely examine their potential to constrain the basic stellar parameters, we show $\amphe$, $\delhe$ and $\tauhe$ in Figures~\ref{fig:fig4} and \ref{fig:fig5} and colour-code model points according to their $M$, $[{\rm Fe}/{\rm H}]_i$, $Y_i$, $t_{\rm age}$ and $\alpha_{\rm MLT}$. 

\begin{figure}
	\includegraphics[width=\columnwidth]{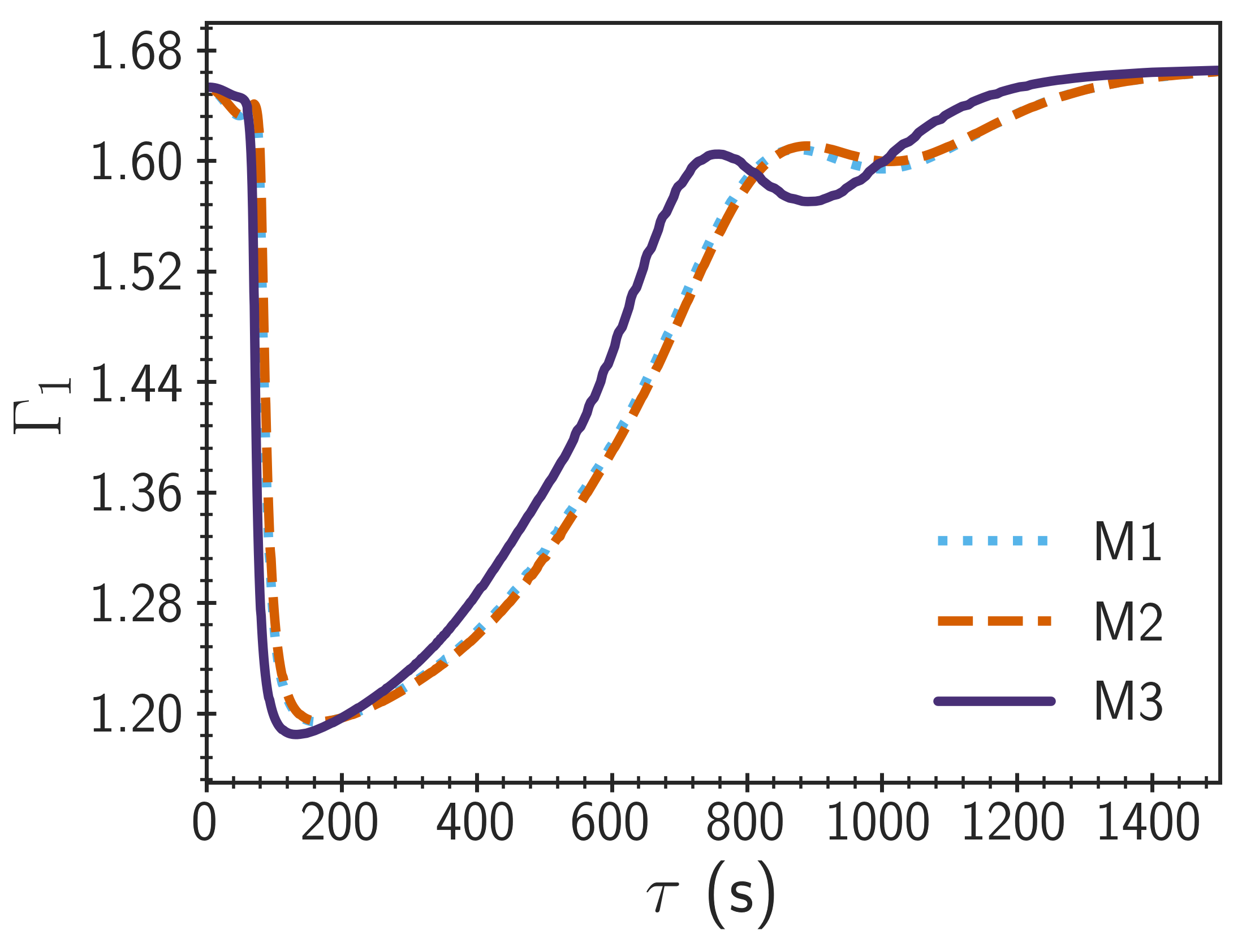}
    \caption{First adiabatic index as a function of acoustic depth. The He glitch parameters for the models M1 ($M = 0.919$ M$_\odot$, $Y_i = 0.249$, $[{\rm Fe}/{\rm H}]_i = 0.088$ dex, $\alpha_{\rm MLT} = 2.044$, $f_{\rm OV} = 0.027$, $t_{\rm age} = 15.875$ Gyr and $T_{\rm eff} = 5575.432$ K) and M2 ($M = 0.911$ M$_\odot$, $Y_i = 0.238$, $[{\rm Fe}/{\rm H}]_i = -0.001$ dex, $\alpha_{\rm MLT} = 2.058$, $f_{\rm OV} = 0.017$, $t_{\rm age} = 16.123$ Gyr and $T_{\rm eff} = 5624.995$ K) do not fall on the trends shown in Figures~\ref{fig:fig4} and \ref{fig:fig5}, whereas for the model M3 (the best-fitting model for 16 Cyg A), they do.}
    \label{fig:fig6}
\end{figure}

It is interesting to note in Figure~\ref{fig:fig4} that $\amphe$ decreases as a function of $\tauhe$ in the left panels (i.e. they are anti-correlated). Moreover, it increases on average with mass as can be noted by observing the colour gradient in the top left panel along the diagonal trend. This is expected because the peak in $\Gamma_1$ between the two stages of helium ionization becomes more prominent for higher masses \citep[see e.g.][]{verm14b,farn19,houd21}. On the other hand, $\delhe$ shows a more complex behaviour in the right panels of Figure~\ref{fig:fig4}; it increases with $\tauhe$ for smaller values but reaches a peak value at about 1000~s and starts to decrease. Furthermore, it is interesting to observe two sequences of models across the full $\tauhe$ range in the top right panel: one with lower values of mass and the other with higher values. The two sequences intersect each other at about a value of $\delhe = 100$~s. In the middle panels, we do not observe any particularly interesting dependence of the He glitch parameters on $[{\rm Fe}/{\rm H}]_i$, which is in any case expected to be well constrained by the measured surface metallicity. In the bottom left panel, we can clearly see that the thickness of the trend is mainly a result of the scatter in $Y_i$, and therefore these observables can constrain $Y_i$ in an effective manner. This is also anticipated because $\amphe$ is known to constrain the surface helium abundance. Although it may appear that the observed point could move significantly in the horizontal direction because of the large corresponding errorbar and cover large $Y_i$ variation, this is not entirely true. The relatively large errorbar on $\tauhe$ is a result of its anti-correlation with $\amphe$ (see Figure~\ref{fig:fig2}), and therefore the observed point is expected to move at an angle (not horizontally) as shown by the confidence ellipse. It is interesting to note that the color of the two sequences get flipped in the bottom right panel of Figure~\ref{fig:fig4} because of the well-known anti-correlation between $M$ and $Y_i$ \citep[see e.g.][]{metc09,lebr14,verm16}. Furthermore, the observed anti-correlation between $M$ and $\amphe$ at a given $\tauhe$ in the top left panel is also a result of this anti-correlation. 

It should be noted that the observed points in Figure~\ref{fig:fig4} fall systematically on the right side of the models, i.e. the model $\tauhe$ values are smaller than the observed value. We wish to point out that this is not due to our choice of the parameter space for the grid nor because of our specific selection of models for calculating the likelihood. Since both the observed and model $\tauhe$ are obtained from the corresponding oscillation frequencies (without making any reference to the surface), we cannot firmly say if this offset is related to the differences in the locations of the acoustic surface. As we will see in Section~\ref{sec:surface_effect}, the observed $\tauhe$ is systematically larger than the corresponding best-fitting model value for all the stars; this small bias is likely a result of the surface effect perturbing the sound speed stratification in the outermost layers and/or the location of the acoustic surface in the models.   

The age trend in the top panels of Figure~\ref{fig:fig5} can be understood in terms of the He glitch parameters' mass dependence (note the reverted color gradients in the top panels of Figure~\ref{fig:fig4}). It is unlikely that the outer layer properties -- such as $\amphe$, $\delhe$ and $\tauhe$ -- have any direct information about the evolutionary changes taking place in the stellar core. The smoother colour gradient for the age is a result of its large relative variation compared to the mass. The dependence of the He glitch parameters on $\alpha_{\rm MLT}$ is very interesting, particularly in the bottom right panel. It shows that $\delhe$ and $\tauhe$ are positively correlated with $\alpha_{\rm MLT}$, and hence our approach can provide a reliable estimate of this stellar parameter as well. 

\begin{table*}
	\centering
	\caption{The properties of the best-fitting models and the corresponding $\chi^2$ values for all the stars in our sample. In the last column, we have also provided the reduced chi-square, $\chi^2_r = \chi^2/(N - f)$, where $N$ and $f$ are the number of observables and free parameters, respectively. The given values for the large frequency separation have been corrected for the surface effect. KIC~12069424 and 12069449 are 16~Cyg~A \& B, respectively.}
	\label{tab:tab2}
	\begin{tabular}{ccccccccc}
		\hline
		KIC & $\Teff$ (K) & $\FeH$ (dex) & $\Dnu$ ($\mu$Hz) & $\amphe$ ($\mu$Hz) & $\delhe$ (s) & $\tauhe$ (s) & $\chi^2$ & $\chi^2_r = \chi^2/(N - f)$\\
		\hline
		3427720  & 5960 & -0.025 & 119.86 & 0.572 & 87.7  & 663.2 & 31.5  & 1.6\\
		6106415  & 5974 & 0.042  & 103.97 & 0.578 & 97.3  & 826.1 & 36.0  & 1.3\\
		8379927  & 6064 & 0.002  & 120.40 & 0.686 & 77.6  & 650.9 & 54.3  & 1.9\\
		9139151  & 6251 & 0.022  & 117.26 & 0.688 & 80.7  & 676.8 & 37.9  & 1.9\\
		10644253 & 6025 & 0.118  & 122.83 & 0.812 & 73.8  & 631.2 & 18.5  & 1.0\\		
		12069424 & 5893 & 0.151  & 103.45 & 0.494 & 109.1 & 885.9 & 63.3  & 2.1\\
		12069449 & 5903 & 0.094  & 117.18 & 0.459 & 97.9  & 768.3 & 117.0 & 4.2\\
		\hline
	\end{tabular}
\end{table*}

There is a small fraction of models (144 out of a total of 56453 models with calculated likelihood) for which the He glitch parameters have arbitrary values and do not fall on the trends shown in Figures~\ref{fig:fig4} and \ref{fig:fig5}. It turns out that these models have low $\Teff$ (< 5700 K) and have either $M$ or $Y_i$ or both close to the grid lower limits (0.9 M$_\odot$ and 0.23, respectively). In Figure~\ref{fig:fig6}, we show $\Gamma_1$ profiles for two such arbitrarily chosen models, M1 and M2, along with the profile for the "best-fitting" model, M3, of 16 Cyg A. Clearly, models M1 and M2 have smaller and -- more importantly -- wider peak between the two stages of helium ionization compared to the best-fitting model M3. The wider peak (or larger $\delhe$) implies faster decay of the amplitude of the He glitch signature as a function of frequency (see Eq.~\ref{eq:glh_fq}), and gives rise to weaker He signature in a given frequency range. In other words, the smaller and wider $\Gamma_1$ peaks lead to much weaker He glitch signatures in such extreme models, making it difficult to reliably detect the He signature and determine the associated parameters. Since these models are reasonably far from 16 Cyg A and lie at the edge of the grid, they should not systematically affect our results.

\begin{figure*}
	\includegraphics[width=\textwidth]{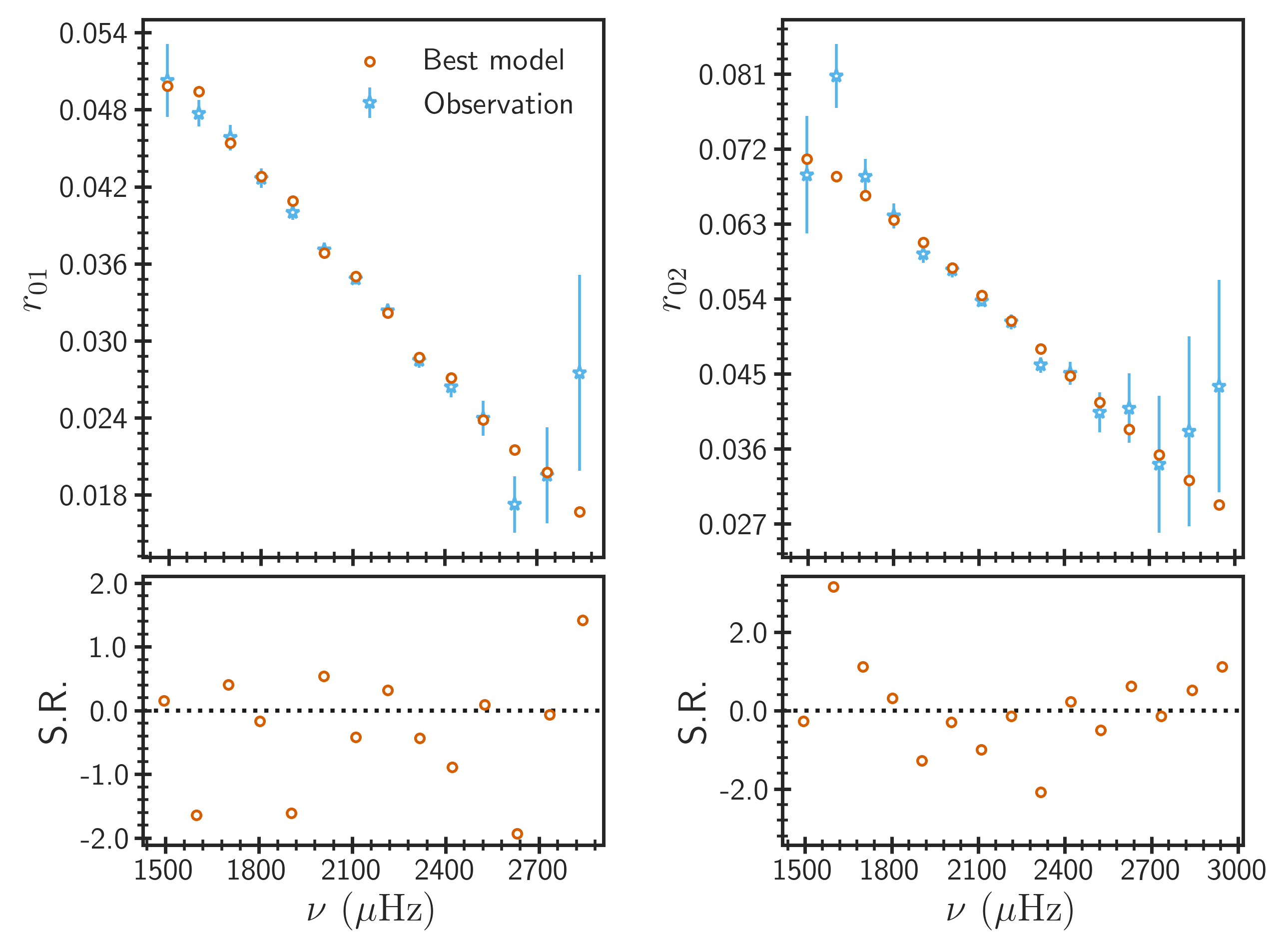}
    \caption{Ratios $r_{01}$ (top left) and $r_{02}$ (top right) for 16~Cyg~A as a function of the observed frequency. The star symbols with errorbar represent the observed data, while the open circles show the best-fitting model values interpolated at the observed frequencies. The bottom panels highlight differences between the observed and model values in the unit of the corresponding observational error (the so-called standardized residual).}
    \label{fig:fig7}
\end{figure*}

We list the properties of the best-fitting model for 16 Cyg A along with the corresponding $\chi^2$ (as defined in Eq.~\ref{eq:chi2}) and $\chi^2_r$ values in Table~\ref{tab:tab2}. Clearly, the model reproduces all the observables well within $2\sigma$. We compare the best-fitting model frequency ratios with the corresponding observed values in Figure~\ref{fig:fig7}. As we can see in the left panels, all the $r_{01}$ ratios agree within $2\sigma$ (with most data points lying within $1\sigma$), and the standardized residuals are approximately randomly distributed around zero without showing any significant trend. In the right panels, we note the same for $r_{02}$, except the ratio at the second lowest frequency for which the model value deviates by about $3\sigma$. To investigate the potential source of this discrepancy, we compare the curvature of the observed and best-fitting model frequency ridges in the \'Echelle diagram (see Figure~\ref{fig:fig2a}) to identify the observed modes that do not follow the curvature of the corresponding model ridges. We find one quadrupole mode with frequency $1590.37\pm0.39$ $\mu$Hz. There are three additional reasons to be suspicious about this observed mode frequency: (1) it lies at the (lower) edge of the power spectrum where the signal-to-noise of the data for all the modes including radial and dipole is generally low; (2) coincidentally, it is a quadrupole mode for which the data has even lower signal-to-noise than the radial and dipole modes; and (3) this mode is involved in the calculation of the discrepant ratio. For these reasons, we suspect that this observed mode frequency may be an outlier and could potentially be the source of the above discrepancy. 

\begin{table*}
	\centering
	\caption{The inferred stellar properties including mass, radius and age for all the stars in our sample. KIC 12069424 and 12069449 are 16 Cyg A \& B, respectively.}
	\label{tab:tab3}
	\begin{tabular}{cccccccc}
		\hline
		KIC & $M$ (${\rm M}_\odot$) & $R$ (${\rm R}_\odot$) & $Y_s$ & $Y_i$ & $[{\rm Fe}/{\rm H}]_i$ (dex) & $\alpha_{\rm MLT}$ & Age (Myr)\\
		\hline
		3427720  & $1.153_{-0.014}^{+0.041}$ & $1.132_{-0.005}^{+0.013}$ & $0.214_{-0.003}^{+0.016}$ & $0.237_{-0.004}^{+0.011}$ & $0.053_{-0.031}^{+0.063}$ & $1.706_{-0.080}^{+0.085}$ & $2351_{-132}^{+113}$\\		
		6106415  & $1.079_{-0.014}^{+0.033}$ & $1.222_{-0.006}^{+0.012}$ & $0.235_{-0.008}^{+0.008}$ & $0.284_{-0.011}^{+0.009}$ & $0.138_{-0.039}^{+0.029}$ & $1.694_{-0.007}^{+0.124}$ & $4688_{-283}^{+60}$\\
		8379927  & $1.126_{-0.022}^{+0.044}$ & $1.124_{-0.009}^{+0.014}$ & $0.251_{-0.010}^{+0.011}$ & $0.275_{-0.015}^{+0.010}$ & $0.102_{-0.051}^{+0.037}$ & $1.698_{-0.136}^{+0.153}$ & $1644_{-101}^{+102}$\\
		9139151  & $1.184_{-0.014}^{+0.010}$ & $1.160_{-0.006}^{+0.004}$ & $0.226_{-0.010}^{+0.014}$ & $0.252_{-0.010}^{+0.013}$ & $0.059_{-0.070}^{+0.060}$ & $1.932_{-0.063}^{+0.108}$ & $2031_{-71}^{+89}$\\
		10644253 & $1.148_{-0.044}^{+0.034}$ & $1.112_{-0.016}^{+0.013}$ & $0.260_{-0.016}^{+0.012}$ & $0.275_{-0.017}^{+0.015}$ & $0.125_{-0.088}^{+0.095}$ & $1.666_{-0.103}^{+0.117}$ & $1194_{-157}^{+145}$\\		
		12069424 & $1.111_{-0.004}^{+0.005}$ & $1.243_{-0.005}^{+0.002}$ & $0.219_{-0.001}^{+0.002}$ & $0.261_{-0.005}^{+0.000}$ & $0.238_{-0.044}^{+0.000}$ & $1.966_{-0.010}^{+0.030}$ & $6719_{-36}^{+236}$\\
		12069449 & $1.067_{-0.061}^{+0.000}$ & $1.126_{-0.027}^{+0.002}$ & $0.224_{-0.000}^{+0.010}$ & $0.257_{-0.000}^{+0.016}$ & $0.161_{-0.000}^{+0.019}$ & $2.046_{-0.206}^{+0.000}$ & $6546_{-30}^{+697}$\\
		\hline
	\end{tabular}
\end{table*}

We list the inferred properties of 16 Cyg A in Table~\ref{tab:tab3}. As can be seen, the inferred mass, radius and the age agree with the corresponding literature values reasonably well. The determined $Y_i$ is larger than the value obtained by \citet{metc15}, thus reducing the bias. \citet{verm19a} found the surface helium abundance for 16~Cyg~A in the range 0.220--0.246 by calibrating the observed $\amphe$ against stellar models of different $Y_s$. Our determination in the table is consistent with their range. We wish to point out that, although our grid is very dense compared to the grids typically used in the literature \citep[for instance, it is about 10 times denser than the grid used in][]{cunh21}, it turns out that it is not dense enough to model the data of quality similar to that of 16 Cyg A (and also B). This results in highly asymmetric uncertainties on the inferred stellar properties.   

Finally, we wish to formally comment on the goodness of our fit based on the value of $\chi^2$. Note that ideally one should calculate the Bayesian evidence for this purpose, however it is difficult to evaluate the involved integrals over the (Sobol) grid of models. Since we fit in total 35 data points for 16 Cyg A (see Table~\ref{tab:tab1}) with 5 free parameters ($M$, $[{\rm Fe}/{\rm H}]_i$, $Y_i$, age and $\alpha_{\rm MLT}$), we expect a value for $\chi^2$ of about 30 for the best-fitting model assuming observational uncertainties (and covariance matrix) associated with various observables are well estimated. Given that the observed metallicity in Table~\ref{tab:tab1} has very small statistical uncertainty (systematic uncertainty is likely to be larger) and one of the low frequency $r_{02}$ ratios fall off the trend in Figure~\ref{fig:fig7}, we cannot confidently say whether a value of $\chi^2 = 63.3$ in Table~\ref{tab:tab2} for 16 Cyg A means that the model needs improvement or that the uncertainties are slightly underestimated. We emphasize that such assessment of goodness of fit is typically not possible when the modelers fit the surface-corrected model frequencies (or anything that involves large systematic uncertainty) as their definition of $\chi^2$ with arbitrary weights to different observables makes it difficult to have any clear expectation of its value for the best-fitting model. 

To study the dependence of the inferred stellar properties on the method used for glitch analysis, we repeated the above exercise for 16 Cyg A using Method B (see Section~\ref{sec:methodb}). Figures~\ref{fig:fig3a} and \ref{fig:fig4a} present the same information obtained using this method as contained in Figures~\ref{fig:fig4} and \ref{fig:fig5} (for Method A). In this case, the total number (387) of models falling off the trends is slightly larger. The errorbars on the second differences are larger by a factor of about 2.5 than those on the frequencies \citep{basu94}, while the increase in the amplitude of the He glitch signature depends on its acoustic depth and is typically more modest \citep{verm17}. In other words, the He signature in the second differences gets effectively diminished compared to that in the frequencies, making it more difficult to extract it using Method B. Having said that, the number of such models are still small in comparison to the total number of models for which the likelihood was calculated. Overall, we obtained very similar results using Method B; in particular, the best-fitting model remains the same.
\begin{figure*}
	\includegraphics[width=\textwidth]{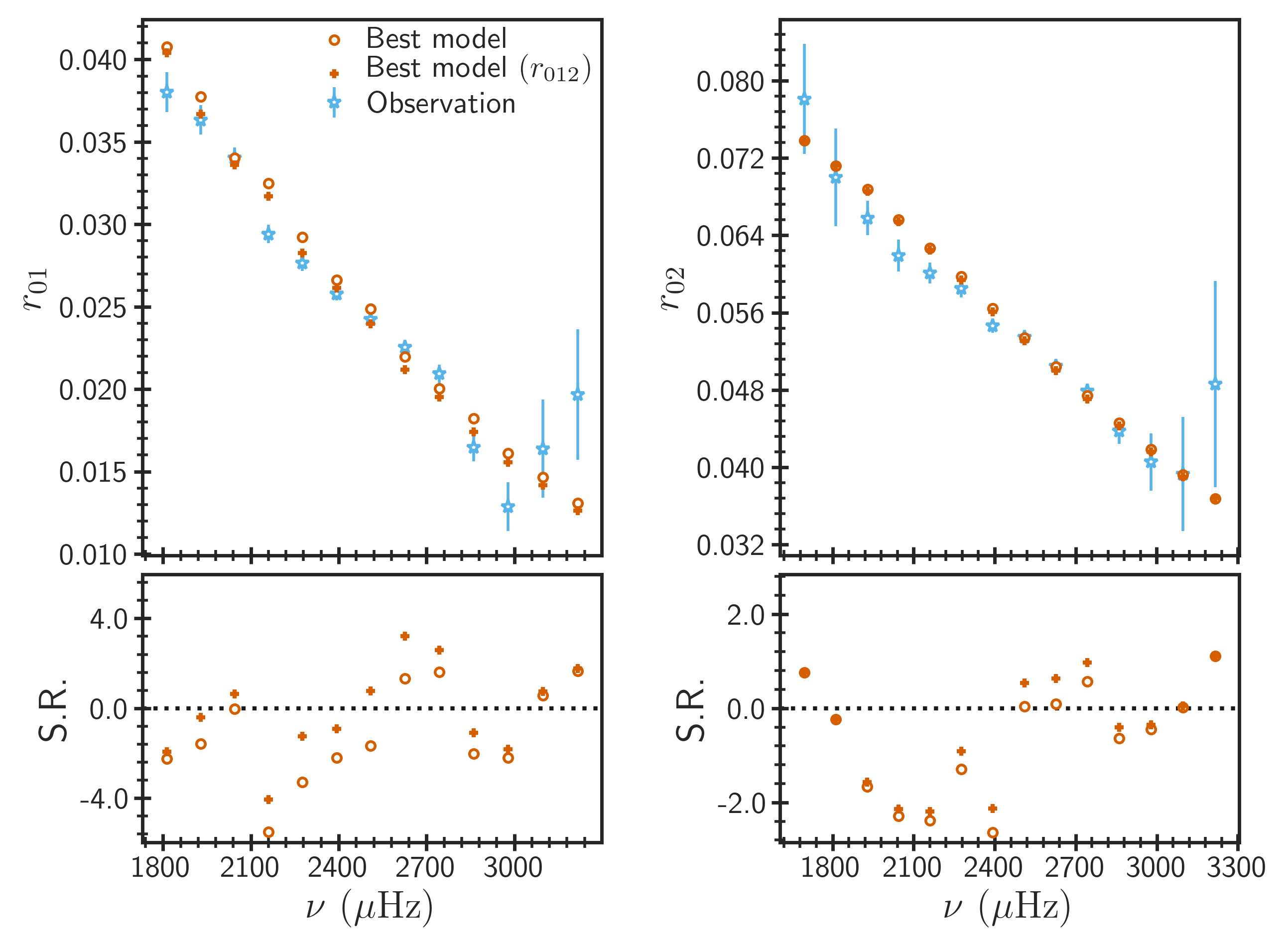}
    \caption{Ratios $r_{01}$ (top left) and $r_{02}$ (top right) for 16 Cyg B as a function of the observed frequency. The star symbols with errorbar represent the observed data, while the open circles show the reference best-fitting model values interpolated at the observed frequencies. The plus symbols represent another best-fitting model obtained by fitting $r_{012}$ (instead of $\gr$). The bottom panels highlight differences between the observed and model values in the unit of the corresponding observational error (the so-called standardized residual).}
    \label{fig:fig8}
\end{figure*}

\subsection{Modelling a sample of LEGACY stars}
%%%%%%%%%%%%%%%%%%%%%%%%%%%%%%%%%%%%%%%%%%%%%%%
\label{sec:sample}
Within the \Kepler asteroseismic LEGACY project, \citet{silv17} used several different approaches to infer properties of all the 66 stars in the sample. We systematically looked at the inferred masses and the measured metallicities and large frequency separations to identify those stars from the LEGACY sample that fell well within the parameter space covered by our stellar model grid described in Section~\ref{sec:approach}. In addition to 16 Cyg B, we found KIC 3427720, 6106415, 8379927, 9139151 and 10644253 which can be modelled using the same grid. We fitted all the observables listed in Table~\ref{tab:tab1} for these stars using the method illustrated in the previous section using the example of 16 Cyg A. We found that most of the trends in Figures~\ref{fig:fig4} and \ref{fig:fig5} for 16 Cyg A including the behavior of $\amphe$ and $\delhe$ as a function of $\tauhe$ and their relationships with mass, helium abundance and mixing-length are generic, and can be seen for other stars as well. The fraction of models with unreliable He glitch parameters remains small (always $ < 1\%$, and for some stars no such models were found). The properties of the best-fitting models of all the stars together with the corresponding $\chi^2$ values are listed in Table~\ref{tab:tab2}, while the inferred stellar parameters are provided in Table~\ref{tab:tab3}. In the following, we discuss the interesting aspects of the results for individual or certain group of stars.

\subsubsection{16 Cyg B}
%%%%%%%%%%%%%%%%%%%%%%%%
The inferred mass, radius and age listed in Table~\ref{tab:tab3} for 16 Cyg B are consistent with the corresponding literature values, and the surface helium abundance is also in agreement with the range, 0.219--0.255, found by \citet{verm19a}. However, it is interesting to note the large value of $\chi^2$ in Table~\ref{tab:tab2}. This is a result of large (more than $3\sigma$) discrepancies between the observed and best-fitting model $\Teff$, $\delhe$ and ratios (see Figure~\ref{fig:fig8}). The standardized residuals in the figure show clear trends. To test whether this is due to some unknown issues related to glitch analysis, we performed a fit with $r_{012}$ (instead of $\gr$). This results in a best-fitting model that predicts $\Teff = 5807$ K, in reasonable agreement with the observed value, and has lower $\chi^2$ ($= 79.7$). Note that $\chi^2$ reduces not just due to the exclusion of the He glitch parameters but also because of the fact that the new best-fitting model has effective temperature and also the ratios to some extent in better agreement with the observation. As we can see in Figure~\ref{fig:fig8} however, the corresponding standardized residuals are still large and follow the same trend as the reference best-fitting model. Moreover, the new best-fitting model has subprimordial $Y_i$ ($= 0.236$), and the inference suffers from the usual bias when excluding the glitch properties from the fit. Therefore, we conclude that the issue is not related to glitch analysis. The large $\chi^2$ then may either mean that the uncertainties on the observed frequencies are significantly underestimated or our stellar models have detectable shortcomings. Relevant to the former case, it is interesting to note that the observed frequencies and hence ratios for 16 Cyg B are slightly more precise than 16 Cyg A (see Table~\ref{tab:tab1}). Given the fact that signal-to-noise of the seismic data depends on the luminosity-to-mass ratio of the star \citep[see][]{kjel95}, the above is in contrary to our expectation of more precise data for the primary component than the secondary. The unreliable zero uncertainties on the inferred parameters for 16 Cyg B in Table~\ref{tab:tab3} are likely results of the inconsistencies among the observables and the insufficient model grid density (as was also noted for 16 Cyg A).

\subsubsection{KIC 6106415 and 8379927}
%%%%%%%%%%%%%%%%%%%%%%%%%%%%%%%%%%%%%%%
These two stars were also part of the sample of 22 stars studied in \citet{math12} using the Asteroseismic Modelling Portal. Interestingly, they found $Y_i$ for KIC 6106415 and 8379927 to be $0.246\pm0.013$ and $0.234\pm0.032$, respectively, which are consistent with the current determination of the primordial helium abundance. In Table~\ref{tab:tab2}, the values of $\chi^2$ of 36.0 and 54.3 for KIC 6106415 and 8379927 ensure that the observables were fitted reasonably well by the corresponding best-fitting models. As can be seen in Table~\ref{tab:tab3}, we found significantly larger values of $Y_i$ for both stars compared to \citet{math12}, reinforcing our belief that our approach overcomes the biased $Y_i$ problem, and hence provides more accurate determinations of the stellar mass and age. The value of $Y_s$ for KIC 8379927 is consistent with the range, 0.237--0.251, found by \citet{verm19a}, whereas it is slightly outside their range of 0.201--0.222 for KIC 6106415.

\subsubsection{KIC 3427720, 9139151 and 10644253}
%%%%%%%%%%%%%%%%%%%%%%%%%%%%%%%%%%%%%%%%%%%%%%%%%
It should be noted from Table~\ref{tab:tab1} that the precision of the seismic data is poorer and the number of observed modes are smaller for these stars compared to the rest. This limits the precision of their measured He glitch properties (see Table~\ref{tab:tab1}). The $\chi^2$ values (see Table~\ref{tab:tab2}) for all the stars are close to what we expect given the number of observables, ensuring reasonable fits for all of them. The inferred $Y_s$ for all the stars are in good agreement with the corresponding ranges (0.191--0.205, 0.213--0.233 and 0.256--0.270 for KIC 3427720, 9139151 and 10644253, respectively) found by \citet{verm19a}. It is interesting to note that the value of $Y_i$ for KIC 3427720 found in this study and also in \citet{verm19a} is consistent with the primordial value within the uncertainty. Although we cannot rule out the possibility of observing individual solar metallicity stars with substantially subsolar helium abundance however, given that this star has only a slightly subsolar metallicity (see Table~\ref{tab:tab1}) and an age of about 2.4 Gyr (see Table~\ref{tab:tab3}), it is unlikely that KIC 3427720 was born with the primordial helium abundance. It would be interesting to reanalyse this star when we have better quality seismic data and hence higher precision He glitch properties.

\begin{figure}
	\includegraphics[width=\columnwidth]{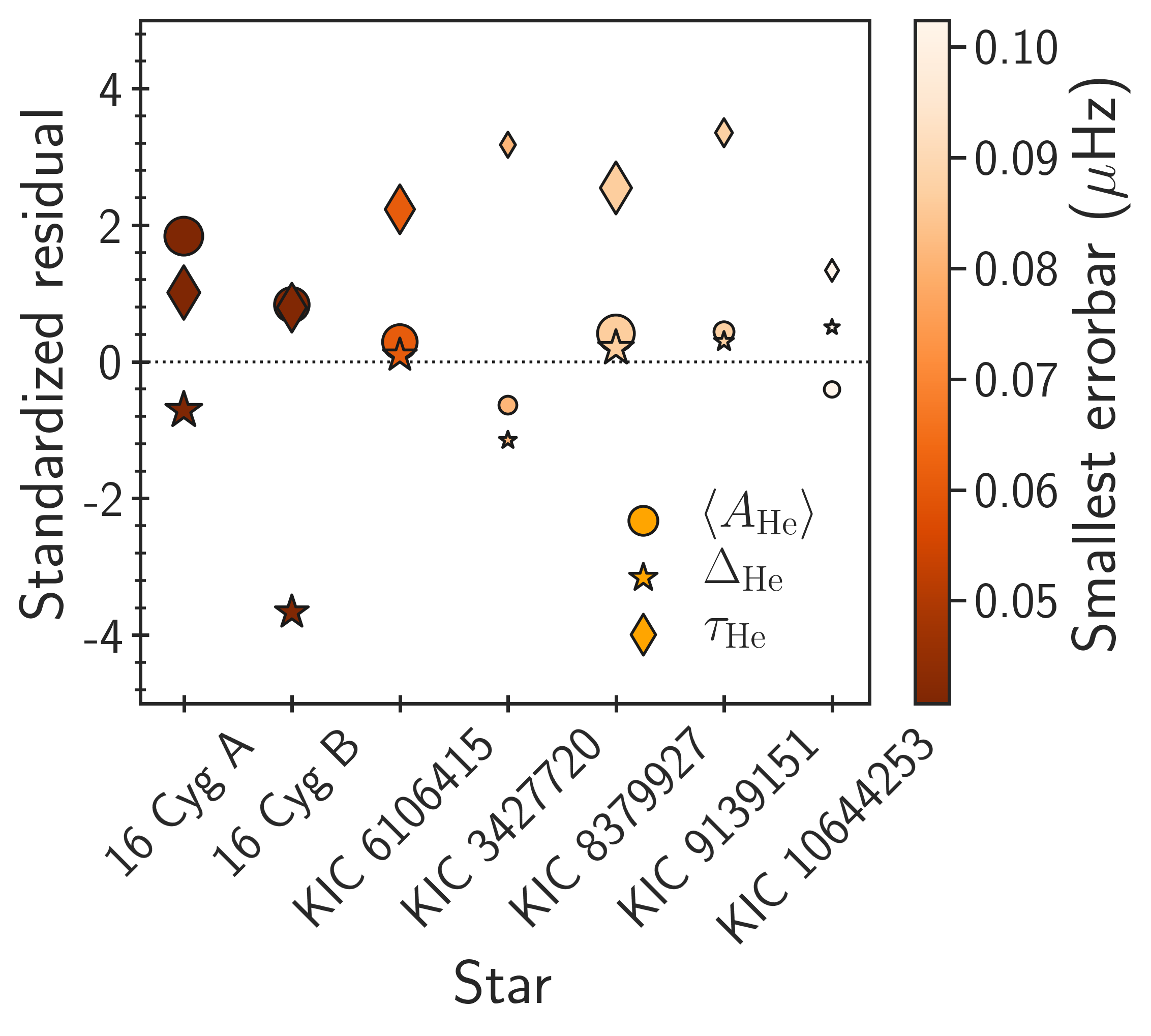}
    \caption{The differences between the observed and best-fitting model He glitch parameters in units of the corresponding observational errors for all the stars analyzed in this study. The different types of points represent different quantities as shown in the legend. The point size has been scaled according to the number of observed modes, while the colour represents the size of the errorbar on the highest precision frequency.}
    \label{fig:fig9}
\end{figure}

\subsection{Impact of the surface effect}
%%%%%%%%%%%%%%%%%%%%%%%%%%%%%%%%%%%%%%%%%
\label{sec:surface_effect}
The surface effect results in a perturbation to the model frequencies that varies slowly (like the smooth component, see Section~\ref{sec:glitch_analysis}) compared to the rapidly varying glitch contributions as a function of the frequency. As a result, this perturbation gets largely filtered out as part of the smooth component, leaving the glitch parameters nearly independent of the surface effect \citep[see appendix B of][]{verm19a}. In Figure~\ref{fig:fig9}, we show the differences between the observed and best-fitting model He glitch parameters in units of the corresponding observational uncertainties to investigate any residual biases due to the surface effect. As we can see, the observed and model $\amphe$ agree well within $1\sigma$ (except for 16~Cyg~A for which it agrees within $2\sigma$), and the residuals are approximately randomly distributed around zero without any significant trend. The same is true for $\delhe$ with the exception this time being 16~Cyg~B. On the other hand, the model $\tauhe$ appears to be systematically lower than the corresponding observed values by on average $2\sigma$. In other words, the stellar models predict on average slightly larger sound speed in the outer layers compared to speed in the stars. Since the data quality can play a role in separating the glitch contributions from the smooth component (which includes the surface term), we scaled the points linearly in Figure~\ref{fig:fig9} with the number of observed modes and colour-coded them with the uncertainty on the highest precision frequency to see if the bias in $\tauhe$ depends on this aspect. The stars with larger and darker points have higher quality seismic data compared to those with smaller and lighter points. The figure indicates that the higher quality data typically leads to smaller bias in $\tauhe$, though there is an exception (KIC 10644253).

\section{Summary and Conclusions}
%%%%%%%%%%%%%%%%%%%%%%%%%%%%%%%%%
\label{sec:conc}
We developed an advanced asteroseismic modelling approach in which the frequency ratios and the He glitch parameters were fitted together using the \BASTA software. The ratios and the He glitch parameters carry complementary information about stellar interior, and help us put tight constraints on it. We used Monte Carlo simulations and the minimum covariance determinant estimator to compute robustly the covariance matrix for these observables, which was subsequently used in the stellar inference problem to properly take into account the correlations. In our approach, we avoided giving any ad-hoc weights to various measured spectroscopic and asteroseismic quantities in contrary to what has previously been done in the literature \cite[see e.g.][]{cunh21}. 

This method was tested on the \Kepler benchmark star, 16~Cyg~A. We showed that the best-fitting model reproduces all the observables reasonably well and we obtained stellar properties including mass, radius and age in agreement with the corresponding literature values. Since the potential of the He glitch properties to constrain the basic stellar parameters (mass, initial metallicity, initial helium abundance, age and mixing-length) has not been explored systematically in the past, we investigated this aspect in Figures~\ref{fig:fig4} and \ref{fig:fig5} in detail. We confirmed the dependence of the average amplitude of He signature on the helium abundance. Moreover, it was interestingly found that both the acoustic depth and width of the He ionization zone correlate positively with the mixing-length parameter and hence they can be used to determine this important quantity reliably. We obtained a larger value of the initial helium abundance for 16~Cyg~A than that found in \citet{metc15}, reducing the bias significantly.

We identified an additional six stars including 16~Cyg~B from the \Kepler asteroseismic LEGACY sample that could be modelled using our grid. Although the fit to the data of 16~Cyg~B resulted in stellar parameters in agreement with the literature values, the best-fitting model did not reproduce the observations satisfactorily. In particular, there were discrepancies at more than $3\sigma$ level between the observation and best-fitting model for the effective temperature, the acoustic width of He ionization zone, and for certain ratios, hinting at issues either in the frequency measurements or in the stellar evolution models. In contrary to our expectation, we noted that the observed frequencies for 16~Cyg~B were slightly more precise than the primary component, 16~Cyg~A. Assuming that there are no issues with the seismic data, it will be interesting to reanalyse 16 Cyg B in the future in more detail with different sets of non-standard input physics, especially those that impact the structure near the base of the convective envelope, to see whether we could reproduce the data better. The modelling of KIC 6106415 and 8379927 provided initial helium abundances systematically larger than the primordial helium abundance and those found by \citet{math12}, reinforcing our belief that this method alleviates substantially the problem of helium abundance being biased towards lower values, and provides more accurate determinations of the stellar mass and age. We note, however, that uncertainties in the input physics can still lead to some extent biases in estimates of the helium abundance \citep[see e.g.][]{farn20}. The inferred surface helium abundances of all the stars modelled in this study were in good agreement with those found in \citet{verm19a} in which they calibrated the average amplitude of He signature against stellar models of different surface helium abundance.

Finally, we demonstrated using our sample of seven test targets that the He glitch parameters are \emph{nearly} independent of the surface effect \citep[see also][]{verm19a,farn19}, more specifically it seems that only model acoustic depth was on average affected at the $2\sigma$ level. As a result, our modelling approach is almost independent of the surface effect, and at same time, has the capability to constrain both the conditions in the stellar core as well as in the envelope. It is computationally expensive but fully automated, making it an ideal seismic tool for inferring precise masses, radii and ages for thousands of stars expected to be observed during the ESA PLATO mission \citep{raue14}. On a modest computer cluster with 100 CPUs, our method will take about a month to analyse 15,000 stars expected to be observed in the PLATO core program (assuming on average 5 CPU hours per star). In this study, we did not focus on the computational efficiency of the method, however it should definitely be possible to improve. For instance, a quick test with 50 instead of 200 random initialization of the fitting parameters (see Section A1) works almost equally well for the models, reducing the time by about a factor of 4.

\section*{Acknowledgements}
%%%%%%%%%%%%%%%%%%%%%%%%%%%
This work was partially supported by the program Unidad de Excelencia Mar\'{i}a de Maeztu CEX2020-001058-M. Funding for the Stellar Astrophysics Centre is provided by The Danish National Research Foundation (Grant agreement no.: DNRF106). We thank the anonymous referee for helpful comments. KV is supported by the Juan de la Cierva fellowship (IJC2019-041344-I). AMS is partially supported by grant PID2019-108709GB-I00 of the MICINN of Spain. VAB-K acknowledges support from VILLUM FONDEN (research grant 10118), the Independent Research Fund Denmark (Research grant 7027-00096B), and the Carlsberg foundation (grant agreement CF19-0649). AS acknowledges support from the European Research Council Consolidator Grant funding scheme (project ASTEROCHRONOMETRY, G.A. n. 772293, http://www.asterochronometry.eu).

\section*{Data Availability}
%%%%%%%%%%%%%%%%%%%%%%%%%%%%
The data underlying this article will be shared on reasonable request to the corresponding author.

% References
%\bibliographystyle{mnras}
%\bibliography{references}

% Appendices
\appendix

\section{The methods for glitch analysis}
%%%%%%%%%%%%%%%%%%%%%%%%%%%%%%%%%%%%%%%%%
\label{sec:glitch_analysis}
There are two popular approaches to study signatures of the He and CZ glitches. In the first approach, one tries to extract the glitch signatures directly from the oscillation frequencies \citep[see e.g.][]{mont94,mont98,mont00,verm14a}, while in the second, from the second differences of frequencies with respect to the radial order, $\delta^2\nu_{n,l} = \nu_{n-1,l} - 2\nu_{n,l} + \nu_{n+1,l}$ \citep[see e.g.][]{goug90a,basu94,basu04,verm14a}. The \GlitchPy code provides options for both approaches; i.e. it can be used for fitting frequencies as well as second differences. These are described below in detail as Methods A and B. 

\begin{figure}
	\includegraphics[width=\columnwidth]{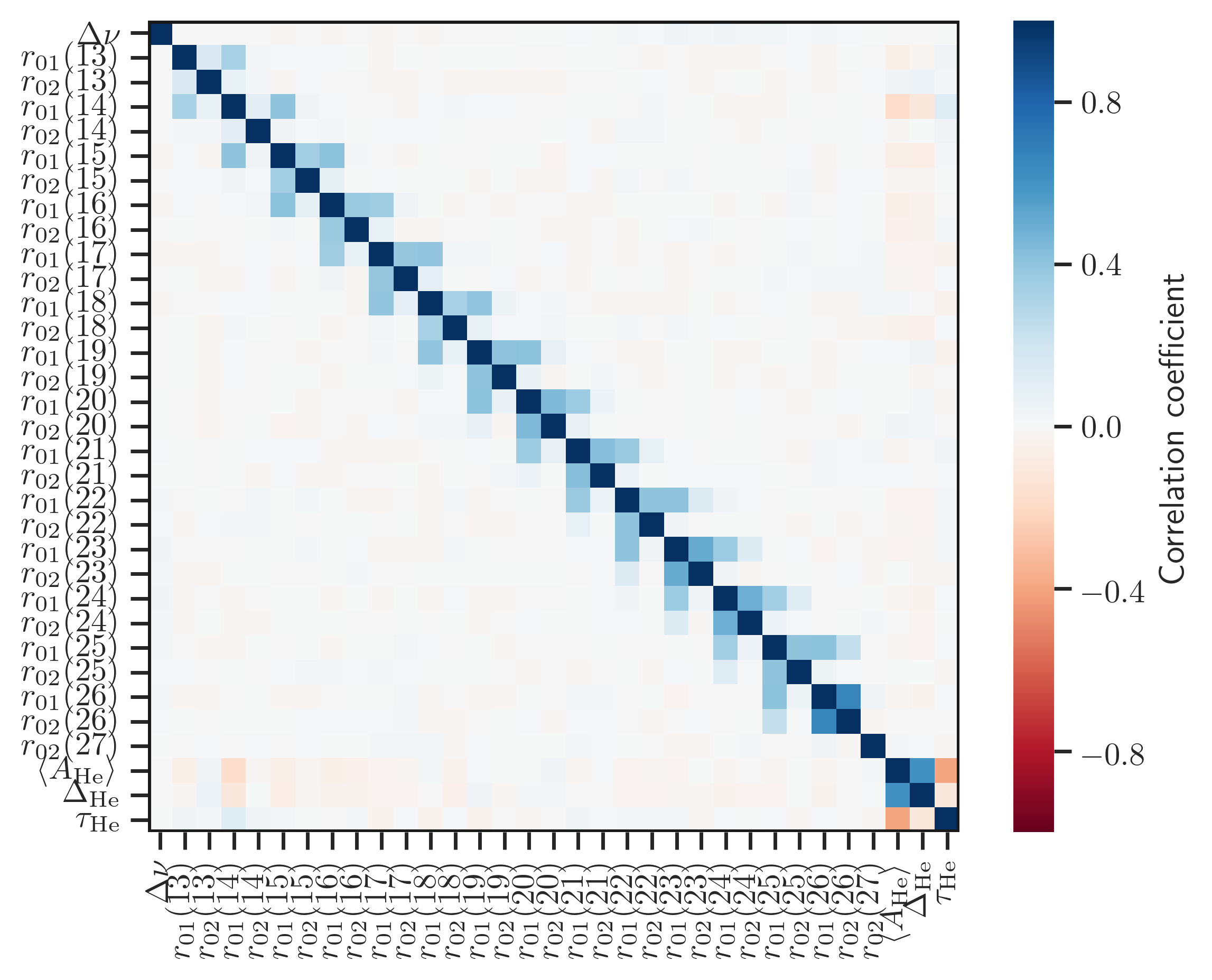}
    \caption{Same as Figure~\ref{fig:fig2}, except it also includes the large frequency separation.}
    \label{fig:fig1a}
\end{figure}

\subsection{Method A: fitting frequencies directly}
%%%%%%%%%%%%%%%%%%%%%%%%%%%%%%%%%%%%%%%%%%%%%%%%%%%
\label{sec:methoda}
We model the smooth component of frequency arising from the smooth structure of the star using a $l$-dependent fourth degree polynomial in $n$,
\begin{equation}
\nu_{\rm smooth}(n,l) = \sum_{k = 0}^4 b_k(l) n^k.
\label{eq:smooth}
\end{equation}
The choice of this particular functional form for the smooth component is partly inspired from the asymptotic theory of stellar oscillations \citep{tass80} and partly from our knowledge about the dependence of the surface term on frequency \citep{kjel08}. Strictly speaking, the above functional form is neither fully consistent with the higher order terms in the asymptotic expansion (as also pointed out by \citet{houd21} in the context of the second difference method) nor with the surface correction proposed by \citet{ball14}. Therefore, this should more be seen as a simple function with large enough degrees of freedom to model everything except the glitch contributions. We emphasize that this function should neither be too simple to adequately describe the smooth component, nor too complex to interfere with the glitch contributions. This is achieved by adopting the above reasonably complex function and subsequently tuning its complexity to a desired level by using regularization in the optimization (see below). We determine the coefficients $b_k(l)$ together with the glitch parameters by fitting the oscillation frequencies to the function,
\begin{eqnarray}
f(n,l) &=& \nu_{\rm smooth} + A_{\rm He} \nu e^{-8\pi^2\delhe^2\nu^2} \sin(4\pi\tauhe\nu + \psi_{\rm He})\nonumber\\
&+& \frac{A_{\rm CZ}}{\nu^2} \sin(4\pi\taucz\nu + \psi_{\rm CZ}),
\label{fitting_function_a}
\end{eqnarray}
where the first term on right hand side represents the contribution to the frequency arising from the background smooth structure, while the second and third terms stand in for the contributions arising from the He and CZ glitches, respectively. The parameters $A_{\rm He}$ and $A_{\rm CZ}$ represent the amplitudes of oscillatory signatures, $\delhe$ the decay rate of the He signature as a function of frequency, $\tauhe$ and $\taucz$ measure the periods of oscillatory signatures, and $\psi_{\rm He}$ and $\psi_{\rm CZ}$ represent the phases.

The fitting is accomplished by minimizing the cost function,
\begin{equation}
\chi_{\rm A}^2 = \sum_{n, l} \left[\frac{\nu_{n,l} - f(n,l)}{\sigma_{n,l}}\right]^2 + \lambda_{\rm A}^2 \sum_{n,l}\left[\frac{d^3\nu_{\rm smooth}}{dn^3}\right]^2,
\label{eq:chi2_a}
\end{equation}
where the first and second terms on right hand side are the usual weighted chi-square ($\sigma_{n,l}$ being the uncertainty on $\nu_{n,l}$) and third derivative regularization, respectively. To determine the regularization parameter, $\lambda_{\rm A}$, we increased it gradually in \citet{verm14a} and monitored the uncertainties on the fitted parameters for 16 Cyg A. The uncertainties dropped quickly in the beginning and reached a plateau. We adopted $\lambda_{\rm A} = 7$ which approximately corresponded to the beginning of the transition. The same value calibrated for 16 Cyg A has been used in several other studies analysing different stars \citep[see e.g.][]{verm17,verm19a}. In this study, we use slightly larger value, $\lambda_{\rm A} = 10$, which marginally improves the precision of the inferred glitch properties for stars with low quality seismic data without affecting much those with high quality data. In \GlitchPy, the user has options to choose any meaningful value for the degree of polynomial in Eq.~\ref{eq:smooth} (must be non-negative), and also values for the order of derivative and regularization parameter in Eq.~\ref{eq:chi2_a}. In principle, terms containing a negative power of $n$ in Eq.~\ref{eq:smooth} can be included, however this has not been currently implemented in \GlitchPy. We wish to emphasize that the use of regularization helps us to effectively tune the degree of the polynomial in a continuous manner.

\begin{figure}
	\includegraphics[width=\columnwidth]{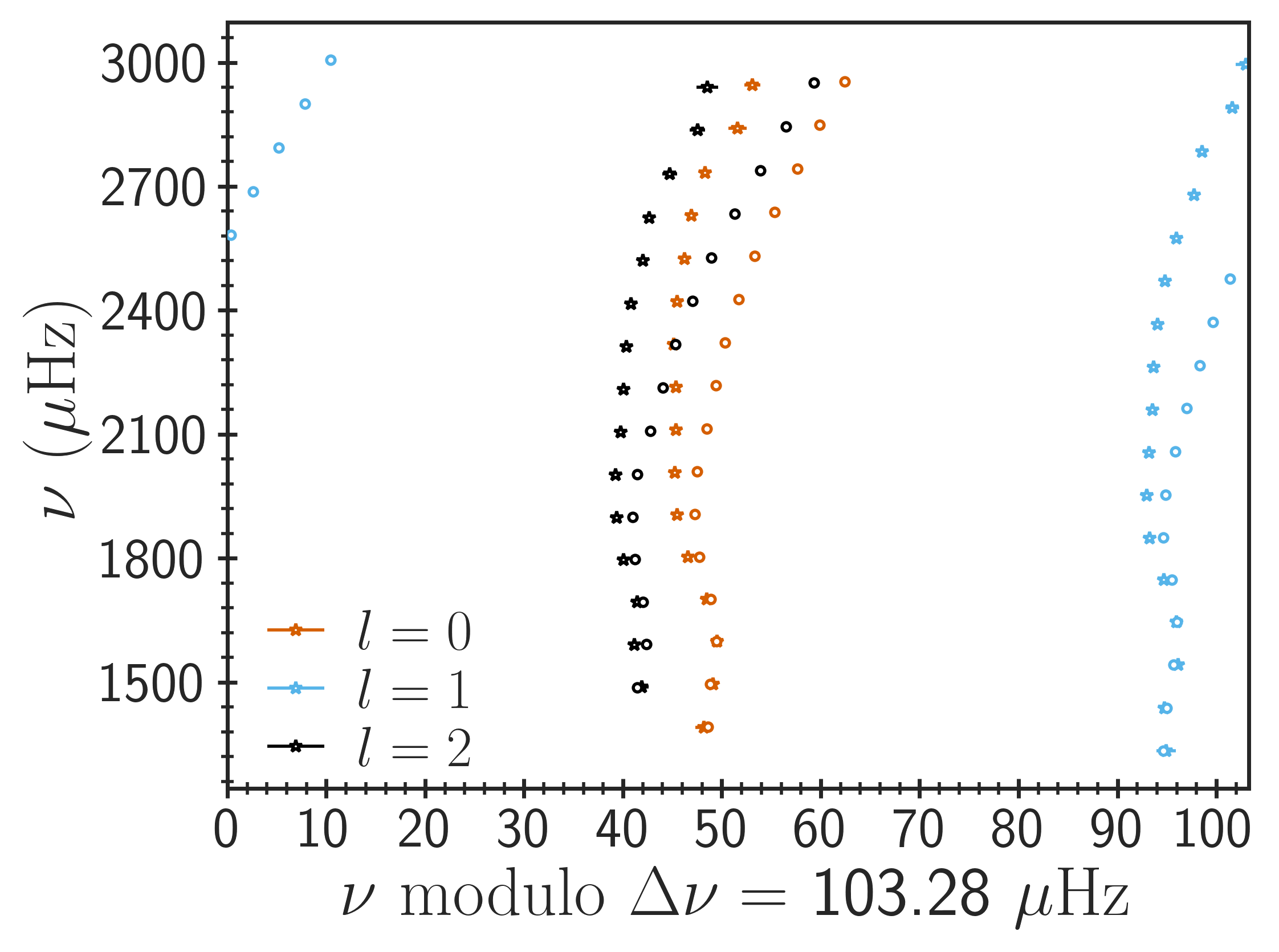}
    \caption{\'Echelle diagram for 16~Cyg~A. The star symbols represent the observed data while the circles show the corresponding best-fitting models. In this case, the model frequencies (used as a reference) were not corrected for the surface effect to clearly show the issue with the observed quadrupole mode of frequency $1590.37\pm0.39$ $\mu$Hz. The colours indicate different harmonic degrees as shown in the legend.}
    \label{fig:fig2a}
\end{figure}
\begin{figure*}
	\includegraphics[width=0.85\textwidth]{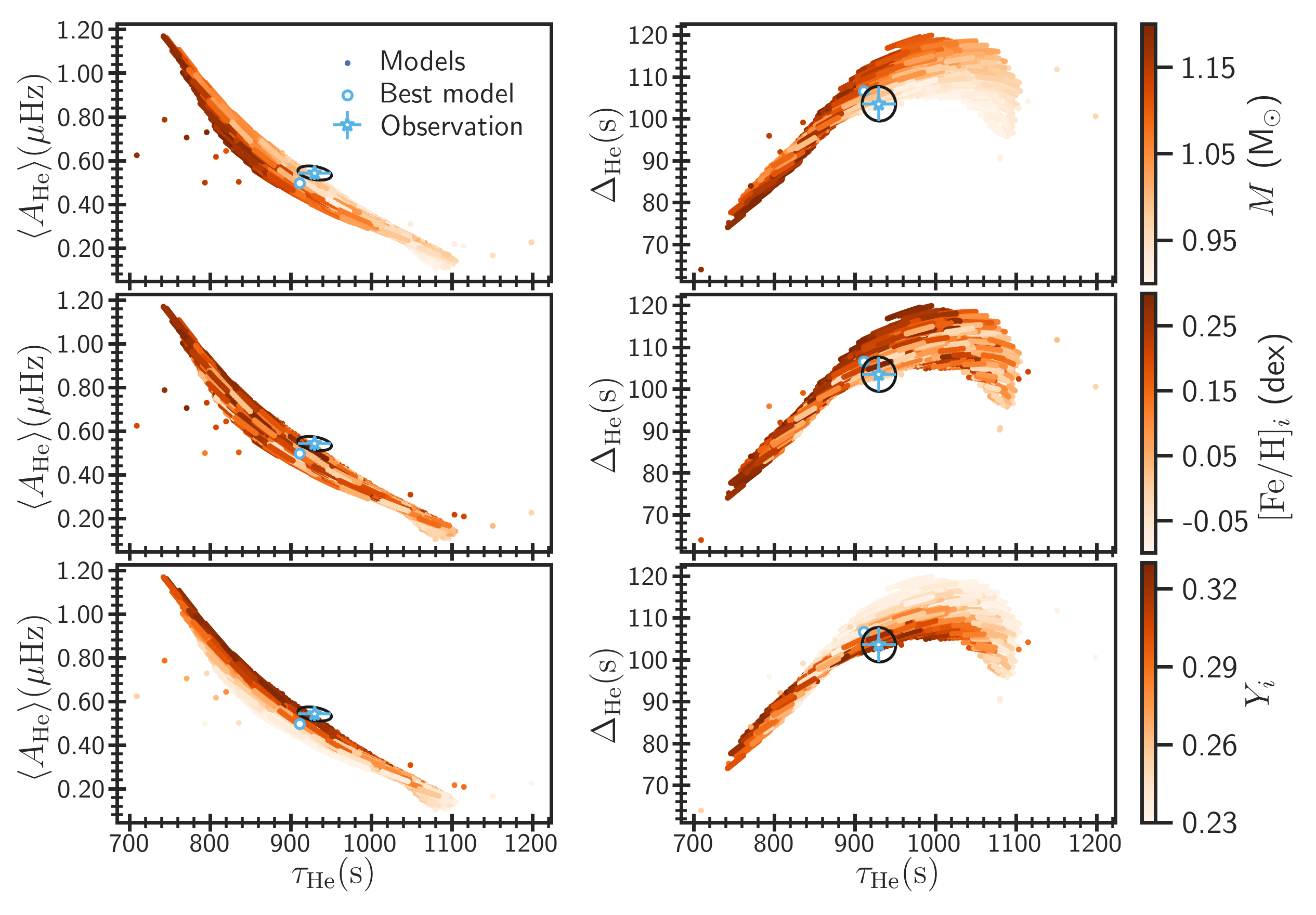}
    \caption{Same as Figure~\ref{fig:fig4}, except the glitch analysis was performed using Method B (instead of A) for both models and the observed data.}
    \label{fig:fig3a}
\end{figure*}
\begin{figure*}
	\includegraphics[width=0.85\textwidth]{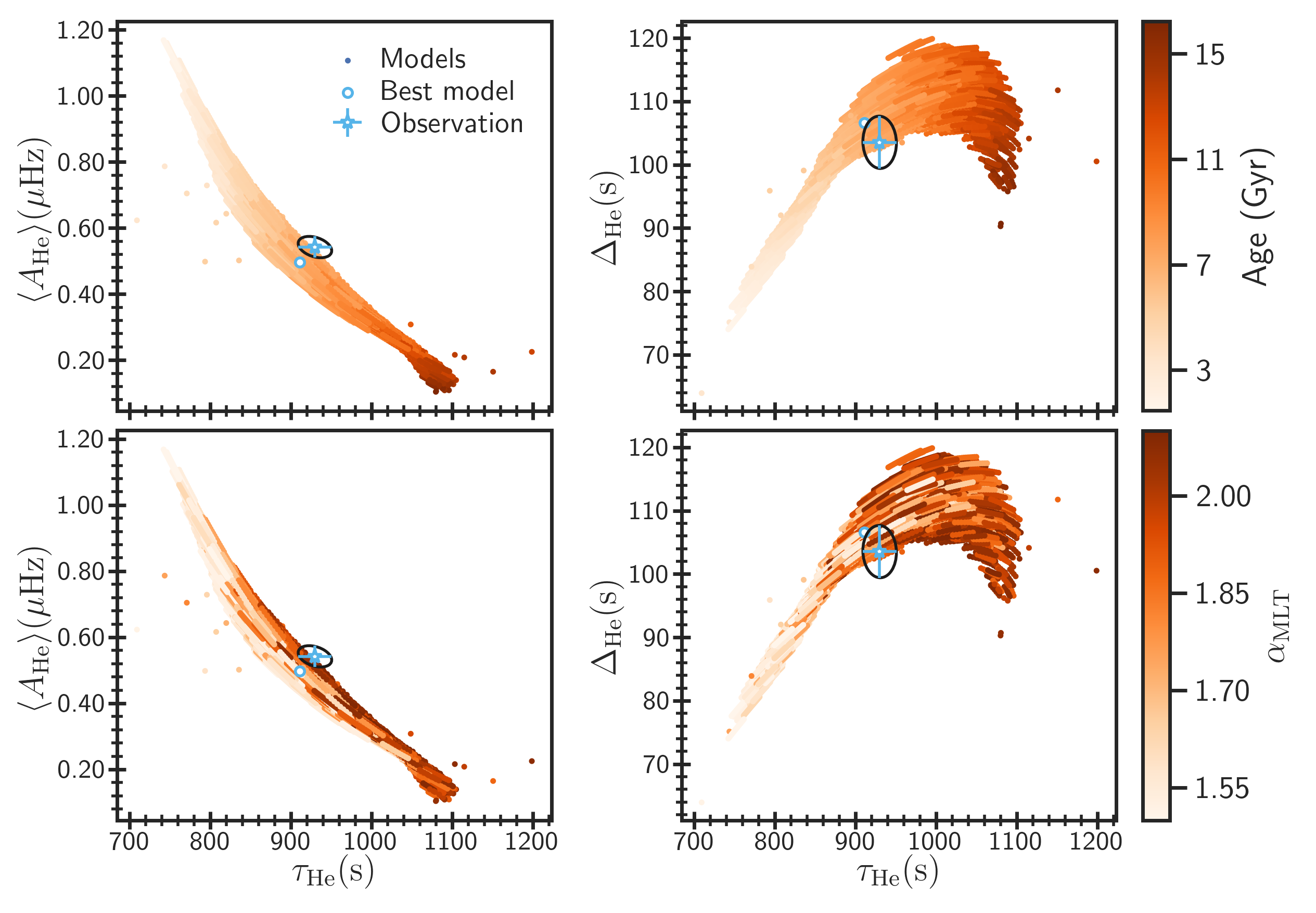}
    \caption{Same as Figure~\ref{fig:fig5}, except the glitch analysis was performed using Method B (instead of A) for both models and the observed data.}
    \label{fig:fig4a}
\end{figure*}
\begin{figure*}
	\includegraphics[width=\textwidth]{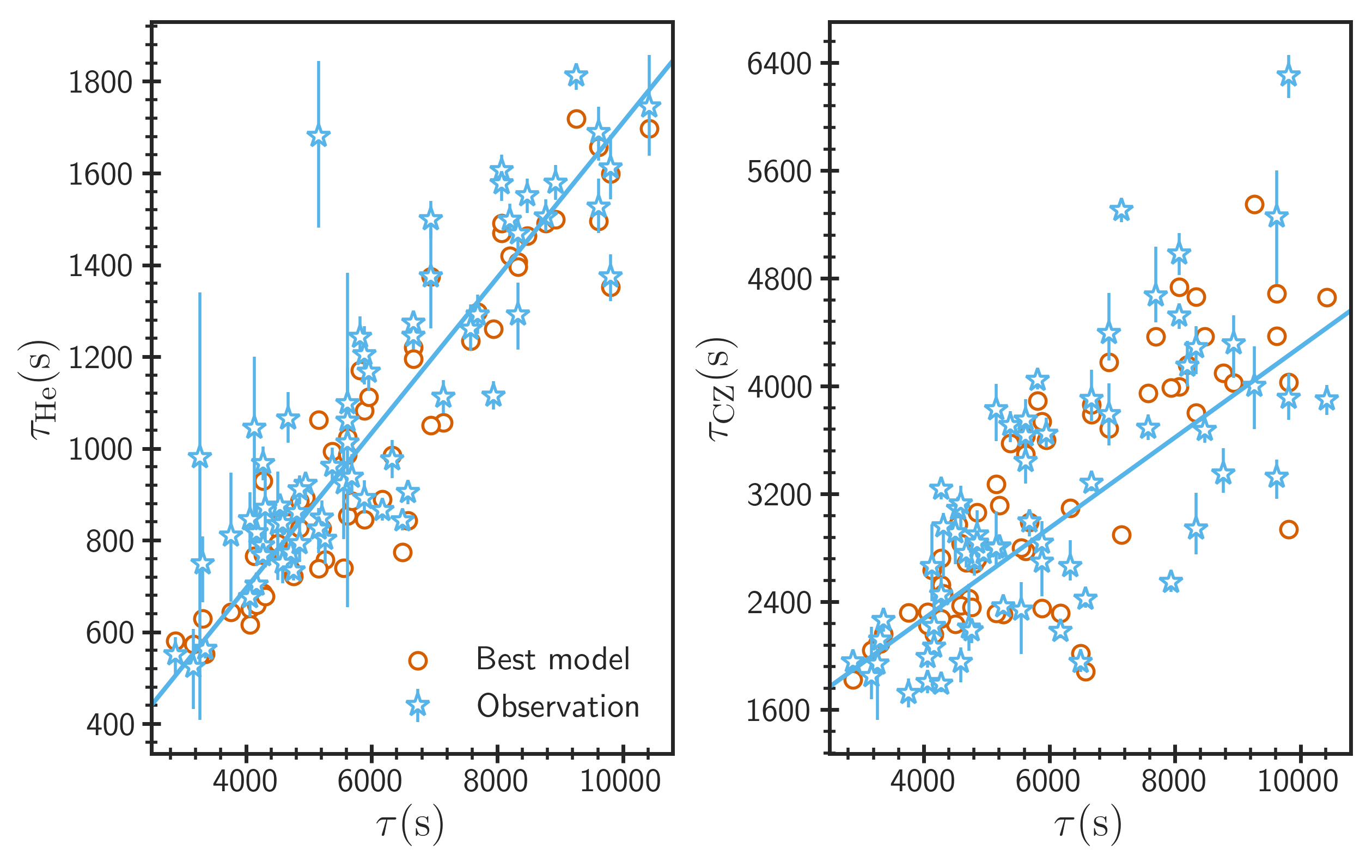}
    \caption{Acoustic depths of the He (left panel) and CZ (right panel) glitches as a function of the acoustic radius of stars in the LEGACY sample. In both panels, the star symbols with errorbar represent the observed stars while the circles show best-fitting models from \citet{verm17}. The straight lines are the weighted linear least-squares fits to the observed data.}
    \label{fig:fig5a}
\end{figure*}

We should note that minimization of $\chi^2_{\rm A}$ w.r.t. $b_k(l)$ and glitch parameters is a non-linear problem, and requires initial guess for the parameters to start the iteration. Given the high-dimension (22 parameters for fitting typically observed radial, dipole and quadrupole mode frequencies), it is possible for the optimizer to get stuck in a local minimum close to the initial guess. To overcome this problem, we perform 200 (user controlled parameter in \GlitchPy) fits by initializing the parameters randomly in a reasonably large space, and accept the fit corresponding to the global minimum. While defining the parameter space, we can safely set the upper and lower limits for both, $\tauhe$ and $\taucz$, to 0 and $(2\Dnu)^{-1}$ (the acoustic radius, $\tau$, of the star), respectively. However, more cleverly defined limits for these two parameters (see Section~\ref{sec:acoustic_depths}) can reduce human intervention significantly, and allow the glitch analysis to be performed in an automated manner. This is particularly useful when we extract glitch properties for tens of thousands of models (for example while stellar modelling, as carried out in this study), as well as when we analyse the observed frequencies for a large sample of stars (specifically relevant for the PLATO mission). We emphasize that the final solution is not limited to the space chosen for the random initialization of the fitting parameters. 

We propagate the statistical uncertainties on the observed oscillation frequencies to the glitch parameters by fitting 10,000 realizations of the data. We derive uncertainties following two different approaches: (1) by computing 16$^{\rm th}$, 50$^{\rm th}$ and 84$^{\rm th}$ percentiles; and (2) by estimating covariance matrix using the minimum covariance determinant estimator \citep{pedr11}. The estimates from both approaches generally agree quite well. While fitting the model frequencies, we use the same set of modes and weights ($1 / \sigma_{n,l}^2$) consistently in Eq.~\ref{eq:chi2_a} as for the observed data. This is the reason for calculating the glitch parameters on the fly during the stellar modelling (and not store them in the grid, which can improve the computational efficiency dramatically).

\subsection{Method B: fitting second differences}
%%%%%%%%%%%%%%%%%%%%%%%%%%%%%%%%%%%%%%%%%%%%%%%%%
\label{sec:methodb}
The underlying approach and the algorithm are similar to Method A. We assume that the smooth component in second differences is independent of $l$ (which is reasonably well justified on grounds of the asymptotic theory and expected form of the surface term), and model it with a quadratic function of frequency, 
\begin{equation}
\delta^2\nu_{\rm smooth} = a_0 + a_1 \nu + a_2 \nu^2,
\label{eq:smooth_b}
\end{equation}
where $a_0$, $a_1$ and $a_2$ are the coefficients. Again, this should be seen as a simple function with large enough degrees of freedom to model everything in the second differences except the glitch contributions. The glitch parameters are determined by fitting the second differences of frequencies to the function,
\begin{eqnarray}
g(n,l) &=& \delta^2\nu_{\rm smooth} + a_{\rm He} \nu e^{-8\pi^2\delhe^2\nu^2} \sin(4\pi\tauhe\nu + \phi_{\rm He})\nonumber\\
&+&\frac{a_{\rm CZ}}{\nu^2} \sin(4\pi\taucz\nu + \phi_{\rm CZ}),
\label{fitting_function_b}
\end{eqnarray}
where the first term on the right hand side represents the contribution to the second difference arising from the background smooth structure, while the second and third terms stand in for the contributions arising from the He and CZ glitches, respectively. The parameters $a_{\rm He}$ and $a_{\rm CZ}$ represent the amplitudes of oscillatory signatures in the second difference (and are different from those in the frequency), $\delhe$ the decay rate of the He signature as a function of frequency, $\tauhe$ and $\taucz$ measure the periods of oscillatory signatures, and $\phi_{\rm He}$ and $\phi_{\rm CZ}$ represent the phases (different from those in the frequency). The amplitudes of the glitch signatures in the second difference can be converted to the corresponding amplitudes in the frequency by dividing them with $4 \sin^2(2 \pi \tau_g \Dnu)$, where $\tau_g$ is the acoustic depth of the glitch \citep[see e.g.][]{basu94}.

We determine $a_0$, $a_1$, $a_2$ and glitch parameters by minimizing the cost function,
\begin{equation}
\chi_{\rm B}^2 = \mathbfit{x}^T \mathbfss{C}^{-1} \mathbfit{x} + \lambda_{\rm B}^2 \sum_{n,l}\left[\frac{d\delta^2\nu_{\rm smooth}}{d\nu}\right]^2,
\label{eq:chi2_b}
\end{equation}
where $\mathbfit{x}$ is a vector containing differences between the observed and model second differences, $\mathbfss{C}$ the analytic covariance matrix for second differences, and $\lambda_{\rm B}$ is the regularization parameter. Previous studies have shown that $\lambda_{\rm B} = 1000$ works well \citep[see e.g.][]{verm19a}. Similar to the Method A, the user has options to choose any meaningful value for the degree of polynomial in Eq.~\ref{eq:smooth_b} (must be non-negative), and also values for the order of derivative and regularization parameter in Eq.~\ref{eq:chi2_b} in \GlitchPy. Furthermore, we find the global minimum and the uncertainties on the glitch parameters in the same way as in Method A. While fitting the model frequencies, we use the same set of second differences and weights ($\mathbfss{C}^{-1}$) consistently in Eq.~\ref{eq:chi2_b} as for the observed data.

\section{Brief descriptions of various figures}
%%%%%%%%%%%%%%%%%%%%%%%%%%%%%%%%%%%%%%%%%%%%%%%
\label{sec:figures}
Figure~\ref{fig:fig1a} presents the full correlation matrix including the large separation, frequency ratios and the He glitch properties for 16 Cyg A (see Section~\ref{sec:covariance} for details). The negligible correlations between $\Dnu$ and the rest of the seismic quantities justifies the separate term in Eq.~\ref{eq:chi2} for $\Dnu$.

Figure~\ref{fig:fig2a} shows the \'Echelle diagram for 16 Cyg A. It demonstrates the potential issue with the observed quadrupole mode with frequency $1590.37\pm0.39$ $\mu$Hz as discussed in Section~\ref{sec:test}. 

Figures~\ref{fig:fig3a} and \ref{fig:fig4a} exhibit the performance of the glitch fitting Method B and highlight the dependence of the glitch properties on the basic stellar parameters. These figures are equivalent to Figures~\ref{fig:fig4} and \ref{fig:fig5} obtained using Method A, respectively (see Section~\ref{sec:test} for details).

\section{Relationships among various acoustic depths}
%%%%%%%%%%%%%%%%%%%%%%%%%%%%%%%%%%%%%%%%%%%%%%%%%%%%%
\label{sec:acoustic_depths}
We use the acoustic depths of the He and CZ glitches for all the stars in the LEGACY sample from \citet{verm17}, and plot them against the acoustic radius (obtained using $\Dnu$) in Figure~\ref{fig:fig5a}. As can be seen, both $\tauhe$ and $\taucz$ are correlated with $\tau$. We fit straight lines to the data to get,
\begin{equation}
    \tauhe = (0.17\pm0.01) \tau + (18\pm59),
\end{equation}
and,
\begin{equation}
    \taucz = (0.34\pm0.05) \tau + (929\pm303).
\end{equation}
These relations provide direct estimates of $\tauhe$ and $\taucz$ for any star with observed value of $\Dnu$, and can be used to define conservative limits as discussed in the previous sections. The estimated $\taucz$ can also be used to differentiate between the real and aliased solutions for the CZ glitch parameters.  

\bsp
\label{lastpage}
\end{document}